\documentclass{aa}
\usepackage{graphicx}
\usepackage{txfonts}
\usepackage{amsmath}	
\usepackage{amssymb}	
\usepackage{multicol}        
\usepackage{bm}		
\usepackage{pdflscape}	
\usepackage{booktabs}
\usepackage{listings}
\usepackage{color}
\usepackage{threeparttable}
\usepackage{multirow}
\usepackage{enumerate}
\usepackage{indentfirst} 
\usepackage{stackengine}
\newcommand\xrowht[2][0]{\addstackgap[.5\dimexpr#2\relax]{\vphantom{#1}}}
\usepackage{booktabs}
\usepackage{subcaption}
\usepackage{lipsum,adjustbox}
\usepackage{arydshln}
\usepackage{autobreak}
\usepackage{enumerate}
\usepackage{txfonts}
\usepackage{ulem}
 \usepackage[bookmarks=false,         
     pdfnewwindow=true,      
     colorlinks=true,    
     linkcolor=blue,     
     citecolor=blue,     
     filecolor=blue,  
     urlcolor=blue,      
final=true
 ]{hyperref}
 \usepackage{etoolbox}
 \usepackage{color}
\usepackage{CJK}
\usepackage{tikz}
\usetikzlibrary{calc, positioning, shapes.geometric}

\newcommand{\TGS}{\texttt{All Galaxies Sample}}
\newcommand{\RSS}{\texttt{Radio Sources Sample}}
\newcommand{\RAS}{\texttt{Radio-excess AGN Sample}}

\begin{document} 
\begin{CJK*}{UTF8}{gbsn}

   \title{Cosmic evolution of radio-excess AGNs in quiescent and star-forming galaxies across $0 < z < 4$}


   \author{Yijun~Wang (王倚君)
          \inst{1,2},
          Tao~Wang (王涛)
          \inst{1,2},
          Daizhong~Liu (刘岱钟)
          \inst{3},
          Mark~T.~Sargent,
          \inst{4}
          Fangyou~Gao (高方宥)
          \inst{1,2},
          David~M.~Alexander
          \inst{5},
          Wiphu~Rujopakarn
          \inst{6,7},
          Luwenjia~Zhou (周陆文嘉)
          \inst{1,2},
          Emanuele~Daddi
          \inst{8},
          Ke~Xu (许可)
          \inst{1,2},
          Kotaro~Kohno
          \inst{9,10},
          Shuowen~Jin (靳硕文)
          \inst{11,12}
          }

   \institute{Department of Astronomy, Nanjing University, Nanjing 210093, China\\
              \email{taowang@nju.edu.cn}
         \and Key Laboratory of Modern Astronomy and Astrophysics (Nanjing University), Ministry of Education, Nanjing 210093, China
         \and Max-Planck-Institut f$\ddot{\rm{u}}$r extraterrestrische Physik, Giessenbachstrasse 1, D-85748, Garching, Germany
         \and International Space Science Institute (ISSI), Hallerstrasse 6, CH-3012 Bern, Switzerland
         \and Centre for Extragalactic Astronomy, Department of Physics, Durham University, South Road, Durham, DH1 3LE, UK
         \and National Astronomical Research Institute of Thailand, Don Kaeo, Mae Rim, Chiang Mai 50180, Thailand
         \and Department of Physics, Faculty of Science, Chulalongkorn University, 254 Phayathai Road, Pathumwan, Bangkok 10330, Thailand
         \and CEA, IRFU, DAp, AIM, Universit$\acute{\rm e}$ Paris-Saclay, Universit$\acute{\rm e}$ Paris Diderot, Sorbonne Paris Cit$\acute{\rm e}$, CNRS, F-91191 Gif-sur-Yvette, France
         \and  Institute of Astronomy, Graduate School of Science, The University of Tokyo, 2-21-1 Osawa, Mitaka, Tokyo 181-0015, Japan
         \and Research Center for the Early Universe, Graduate School of Science, The University of Tokyo, 7-3-1 Hongo, Bunkyo-ku, Tokyo 113-0033, Japan
         \and Cosmic Dawn Center (DAWN), Copenhagen, Denmark
         \and DTU Space, Technical University of Denmark, Elektrovej 327, 2800 Kgs. Lyngby, Denmark
             }

  \abstract
 {Radio-excess active galactic nuclei (radio-AGNs) are essential to our 
understanding of both the physics of black hole (BHs) accretion and the interaction between BHs and host galaxies.
Recent deep and wide radio continuum surveys have made it possible to study radio-AGNs down to lower luminosities
and up to higher redshifts than previous studies, providing new insights into the abundance and physical origin of radio-AGNs.}
{Here we focus on the cosmic evolution, 
physical properties and AGN-host galaxy connections of radio-AGNs 
selected from a total sample of $\sim$ 400,000 galaxies at $0 < z < 4$ in the GOODS-N and COSMOS fields.} 
{Combining the deep radio continuum data with multi-band, de-blended far-infrared and sub-millimeter data, 
we are able to identify 983 radio-AGNs out of the entire galaxy sample 
through radio excess relative to the far-infrared-radio relation.}
{We study the cosmic evolution of 1.4 GHz radio luminosity functions (RLFs) 
for both star-forming galaxies (SFGs) and radio-AGNs, which can be well described by a 
pure luminosity evolution of $L_{\star}\propto (1+z)^{-0.34\times z +3.57}$ and 
a pure density evolution of $\Phi_{\star} \propto (1+z)^{-0.77\times z +2.69}$, respectively. 
We derive the turnover luminosity, above which the number density of radio-AGNs surpasses that of SFGs.
We show that this crossover luminosity increases as increasing redshifts, 
from $10^{22.9}\ \rm{W}\ \rm{Hz}^{-1}$ at $z\sim 0$ to $10^{25.2}\ \rm{W}\ \rm{Hz}^{-1}$ at $z\sim 4$.
At full redshift range of $0 < z < 4$, we further derive the probability ($p_{\rm radio}$) 
of SFGs and quiescent galaxies (QGs) hosting a radio-AGN, 
as a function of stellar mass ($M_{\star}$), radio luminosity ($L_{\rm R}$), and redshift ($z$), 
which yields $p_{\rm radio} \propto (1+z)^{3.08}\ M_{\star}^{1.06}\ L_{\rm R}^{-0.77}$ for SFGs, 
and $p_{\rm radio} \propto (1+z)^{2.47}\ M_{\star}^{1.41}\ L_{\rm R}^{-0.60}$ for QGs, respectively. }
{The quantitative relation for the probabilities of galaxies hosting a radio-AGN 
indicates that radio-AGNs in QGs prefer to reside in more massive galaxies with larger $L_{\rm R}$ than those in SFGs.
The fraction of radio-AGN increases towards higher redshift in both SFGs and QGs, with a more rapid increase in SFGs.
}


   \authorrunning{Y.J. Wang et al.}
   \titlerunning{Cosmic evolution of radio AGNs in SFGs and QGs at $0 < z < 4$}
   \maketitle

%
%
\section{Introduction}
\label{sec:intro}

Active galactic nuclei (AGNs) are believed to play an important role in the growth of 
galaxies through high kinetic 
or radiative power \citep[AGN feedback;][for a general review]{Fiore2017,Matzeu2023,Fabian2012}.
AGNs emit energy across the whole spectrum, 
which can be identified in multiple wavelengths,
including optical \citep{Baldwin1981,Kauffmann2003,Kewley2013}, 
mid-infrared \citep[MIR;][]{Lacy2004,Donley2012}, X-ray \citep[e.g.,][]{Xue2016,Luo2017},
and radio bands \citep[e.g.,][]{DelMoro2013}.
Approaches through optical/MIR/X-ray surveys are sensitive to select AGNs with relatively high Eddington ratio,
while selection through radio bands can find more AGNs with weak nuclear activities
and help to build a more complete AGN sample
\citep{Delvecchio2017,Radcliffe2021}.
Radio AGNs have been found to preferentially reside in
massive quiescent galaxies (QGs) \citep[][for a general review]{Condon1978,Brown2011,Vaddi2016,Ho2008},
and in dense environments \citep{Best2005,Worpel2013,Pasini2022}.
They may make a significant effect on the evolution of their host galaxies and environments through
the mechanical/jet/radio mode feedback \citep[][for reviews]{Fabian2012,Hardcastle2020,Kondapally2023,Magliocchetti2022}.
Constructing a large and complete radio AGN 
sample over a broad luminosity range and a wide redshift range can greatly enlarge our understanding about their abundance and occupation fraction, 
and improve our knowledge about the impact AGN feedback has on host galaxies and their environments.

One effective approach to select radio-AGNs is through selecting objects
with radio emission exceeding that expected from star formation, especially for less luminous radio-AGNs.
In the star formation process, 
IR emission is expected to be correlated with radio emission
due to their mutual origins from activities of massive stars \citep{Condon1992,Dubner2015}, 
so-called IR-radio correlation (IRRC),
which has been established by many observational studies 
\citep[e.g.][]{Helou1985,Condon1992,Yun2001,Appleton2004,Ivison2010,Sargent2010a,Sargent2010b,Delhaize2017,Molnar2021}.
This IRRC can be used to identify radio AGNs
\citep[also be called as radio-excess AGNs;][]{Donley2005,Park2008,DelMoro2013,CalistroRivera2017},
because AGNs may have extra radio emission related to nuclear activities, such as radio jets, AGN-driven outflows, and the innermost accretion disk coronal activities \citep[][for a review]{Panessa2019}.
This method can select many AGNs that are usually missed by 
other established techniques, 
such as optical, X-rays, or MIR surveys \citep{Park2008,DelMoro2013}.

Separating radio-AGNs from SFGs requires both sensitive radio and far-infrared observations, 
which is particularly important for low luminosity radio-AGNs. In the local Universe,
it has been known that SFGs dominate at the low radio luminosity end
while AGNs constitute a significant fraction 
of radio populations at high luminosity end \citep[e.g.,][]{Seymour2008,Condon2019,Franzen2021,Matthews2021}.
At higher redshift, however, large uncertainty remains due to 
the difficulty in simultaneously detecting faint and luminous objects 
with both IR and radio facilities \citep[e.g.,][]{Novak2018}.
The situation has been significantly improved with the accomplishment
of deep and wide radio surveys, e.g., the 3 GHz VLA COSMOS survey 
\citep{Smolcic2017a,Novak2017,Smolcic2017b,Delvecchio2022},
the Low Frequency Array Two-metre Sky Survey (LoTSS) Deep Fields \citep{Best2023,Sabater2021,Tasse2021},
and the MeerKAT International GHz Tiered Extragalactic Exploration (MIGHTEE) Survey \citep{Jarvis2016}.
Moreover, even deeper VLA \citep{Owen2018,Alberts2020} and far-infrared surveys \citep[][Wang et al. in prep.]{Liu2018}
are available in the GOODS fields,
these improvements greatly enlarge the radio sample with more faint objects.
Furthermore, thanks to the detailed de-blended photometry 
for the FIR/submillimeter imaging data 
in the GOODS and COSMOS fields \citep[][Wang et al. in prep.]{Liu2018,Jin2018},
more precise estimates for the IR luminosity are available to
further improve the distinguishment between different radio populations.
Combining the power of these wide and deep surveys 
has great potentials to constrain the abundance and 
physical properties of radio-AGNs down to lower luminosities and up to higher redshifts. 

In addition to the abundances of radio-AGNs, the occupation fraction of radio-AGNs 
in different galaxy populations is also essential to constrain models of 
AGN accretion and feedback. 
While the most luminous radio-AGNs are primarily located in massive quiescent galaxies
in the local universe \citep[e.g.,][]{Peacock1991,Magliocchetti2004,Donoso2010,Worpel2013,Kolwa2019,Dullo2023}, 
it remains unclear whether this is the case at high redshifts \citep[e.g., $z\sim 2$,][]{Malavasi2015}, 
and for the less luminous ones \citep{Uchiyama2022}.
Radio luminosities of AGN may fundamentally depend on the efficiency of gas accretion onto the black hole (BH),
which can be reflected in the BH fundamental plane 
\citep[$L_{\rm R}-L_{\rm X}-M_{\rm BH}$ relation; e.g.,][]{Merloni2003,Bonchi2013,Xie2017,Bariuan2022}.
Many works have systematically investigated the dependence of 
the radio-loud AGN (RL AGN) fraction on AGN radio luminosity,
stellar mass (BH mass), and galaxy types of host galaxies 
especially in the local universe.
RL-AGN fractions are found to increase with stellar mass 
in the local universe \citep[$0.03 < z < 0.3$;][]{Best2005,Sabater2019}
and higher redshift \citep[$\sim 2$;][]{Williams2015}.
This stellar-mass dependence may decrease 
with redshift \citep[from $z \sim 0.3$ to $z \sim 2$;][]{Williams2015,Zhu2023}.
Moreover, RL-AGN fraction may decrease with radio luminosity \citep{Best2005,Sabater2019},
and increase with redshift \citep[][]{Donoso2009,Zheng2022}.
Galaxy colors or galaxy populations have a significant effect on the 
RL-AGN fraction \citep{Janssen2012,Kondapally2022},
which may indicate that supermassive black hole (SMBH) are fuelled by different mechanisms in different galaxy populations \citep{Kondapally2022}.
However, at higher redshift and for fainter radio-AGNs, 
it is still unclear about the dependence of radio-AGN fraction on diverse physical properties of galaxies.
Thanks to the deeper and wider radio surveys \citep[e.g.,][]{Smolcic2017a,Owen2018,Alberts2020,Best2023},
studying the cosmic evolution of both weak and powerful radio-AGNs has become possible, 
which will greatly enrich our understanding about the physical properties of radio-AGNs and their feedback.
Moreover, given that different galaxy populations (e.g., SFGs and QGs) 
may present different associations with AGN feedback
\citep[][for a general review]{Fiore2017,Delvecchio2022,Matzeu2023,Fabian2012},
systematic studies about the radio AGN fractions in different galaxy populations are required
to help us further understand the detailed effects of AGN feedback.
In addition, most of previous studies usually estimated the AGN 
fraction as a function of solely stellar mass, luminosity, or redshift.
To systematically investigate the possible physical properties affecting the radio activities of AGN,
a unified quantitative relation describing radio-AGN fractions as
a function of stellar mass, radio luminosity, and redshift is required.
We may expect this unified relation to serve as an important complement for AGN feedback mode in simulations,
such as Illustris$\textsc{TNG}$ \citep{Weinberger2017,Pillepich2018} and $\textsc{SIMBA}$ \citep{Dave2019},
in the future.

In this work, 
we firstly use the UV/optical/MIR surveys in the
GOODS-N and COSMOS/UltraVISTA fields (totaling 1.55 degree$^2$)
to construct a large galaxy sample at $0.1 < z < 4$ (Section \ref{sec:data}).
Then we cross-match this UV/optical/MIR catalog with
the deep/large radio surveys and de-blended IR luminosity catalogs in these fields (Section \ref{sec:data}).
Next we calculate IR and radio luminosities for individual objects (Section \ref{sec:parameter}),
and use IR-radio-luminosity-ratio to 
select radio-excess AGNs at $0.1 < z < 4$ (Section \ref{sec:REAGNpaper}).
Further, we investigate the cosmic evolution and physical properties 
of radio-excess AGNs through the following two aspects:
(1) constructing radio luminosity functions for SFGs and radio-excess AGNs, repectively,
and studying their evolution with redshift (Section \ref{sec:RLF});
(2) calculating the radio-excess AGN fraction as a function of stellar mass, radio luminosity, and redshift
in different galaxy populations such as SFGs and QGs (Section \ref{sec:Fagntotalsummary}).
The interpretation about our results are discussed in Section \ref{sec:Totdis}.
We summarize our conclusions in Section \ref{sec:summary}.
Throughout this paper, we assume a \cite{Chabrier2003} initial mass function (IMF)
and a flat cosmology with the following parameters:
$\Omega_{\rm m}=0.3$, $\Omega_{\Lambda}=0.7$, and $H_0=70\ \rm{km}\ \rm{s}^{-1}\ \rm{Mpc}^{-1}$.

\section{Data and samples}
\label{sec:data}

Our sample is selected from two fields: 
the Great Observatories Origins Deep Survey North \citep[GOODS-N;][]{Barro2019}
and
the UltraVISTA survey in the Cosmic Evolution Survey \citep[COSMOS;][]{Scoville2007,Weaver2022}.
The GOODS-N survey belonging to the 
deep fields in the Cosmic Assembly Near-infrared Deep Extragalactic Legacy Survey 
\citep[CANDELS; ][]{Grogin2011,Koekemoer2011} 
is suitable to study faint objects in this work.
The GOODS-N field has a region size of 171 arcmin$^2$,
which is not large enough to detect many luminous objects.
Therefore, we also use observational data from the COSMOS/UltraVISTA field 
with a region size of 1.5 degree$^2$.
The COSMOS/UltraVISTA field is within the full COSMOS field that is the largest {\it HST} survey 
imaging a 2 deg$^2$ field \citep{Scoville2007,Weaver2022}.
Here we only use the data from the COSMOS/UltraVISTA field in order to
utilize the de-blended FIR/sub-mm photometry catalog from \cite{Jin2018}.
In addition, the GOODS-N field 
has deeper radio \citep{Owen2018} and 
IR \citep[][]{Liu2018} surveys than the COSMOS field \citep{Smolcic2017a,Jin2018}.
Therefore, summing up the data from these two fields can help us construct a large sample including both faint and luminous objects.
The available multi-wavelength data and total samples from the GOODS-N and COSMOS fields 
in this work are summarized in Figure \ref{fig:flowchart}.
Next we show the detailed matching processes and 
sample selections in each field (the overall flowchart is summarized in Figure \ref{fig:flowchart}).

\subsection{GOODS-N field}
Firstly, we collected UV-optical-MIR data in the GOODS-N field from \cite{Barro2019} (hereafter B19).
This UV-optical-MIR catalog covers the wavelength range between 0.4 and 8 $\mu$m, 
and contains 35,445 sources over 171 arcmin$^2$. We selected 29,267 sources as our {\TGS}
according to the following two criteria: (1) $0.1 < z \leq 4.0$; (2) signal-to-noise ratio (S/N) of F160W band $\geq 5$.

Next, FIR-submillimeter-radio data in the GOODS-N field are derived from \cite{Liu2018} (hereafter L18).
This FIR-submillimeter-radio catalog was established using {\it{Spitzer}} 24 $\mu$m or VLA 20 cm detected sources
as priors for FIR/submm photometry which have potentially significant source confusion.
This ``super-deblended'' photometry method provides more 
accurate estimates for FIR/submm flux of each individual source. 
This catalog contains 3306 sources with photometry at MIR ({\it Spitzer} 16 and 24 $\mu$m), 
FIR ({\it Herschel} 100, 160, 250, 350, and 500 $\mu$m), 
submm (SCUBA2 850 $\mu$m and AzTEC+MAMBO 1160 $\mu$m), and radio (VLA 1.4 GHz) bands.
The 1.4 GHz data are primarily from \cite{Owen2018} 
with rms noise ($\sigma$) in the radio image center of 2.2 $\mu$Jy,
supplemented by \cite{Morrison2010} with $\sigma$ in the radio image center of 3.9 $\mu$Jy beam$^{-1}$.
For sources weaker than 5$\sigma$ detection limit of \cite{Owen2018},
\cite{Liu2018} performed prior-extraction photometry and Monte Carlo simulations on
the radio image of both \cite{Owen2018} and \cite{Morrison2010}.

Then we cross-matched the FIR-submillimeter-radio catalog (L18) 
with the UV-optical-MIR catalog (B19) by a match radius of 1.5 arcsec.
The match radius is defined according to the average angular resolution of 1.4 GHz VLA survey in the GOODS-N field, which is 1.6 arcsec \citep{Owen2018}. 
2584 (of 3306) objects in L18 have UV/optical/MIR counterparts in B19.
The remaining 722 sources in L18 is out of the CANDELS F160W mosaic region
over 171 arcmin$^2$ in the CANDELS/GOODS-N field,
which will not be used in this work.

Finally, among the above-mentioned 2584 sources, we selected 509 sources as our {\RSS}
according to the following two criteria: (1) $0.1 < z \leq 4.0$;
(2) S/N of 1.4 GHz radio flux $\geq 5$.
The {\RSS} have multi-wavelength data (UV-optical-IR-submm-radio bands)
which can be used to estimate various galaxy properties 
(such as 8--1000 $\mu$m IR luminosity $L_{\rm TIR}$ and stellar mass $M_{\star}$) through broadband
spectral-energy-distribution (SED) fitting with \texttt{CIGALE} (see details in Section \ref{sec:LIRM}).

\subsection{COSMOS/UltraVISTA field}
Firstly, UV-optical-MIR catalog in the COSMOS/UltraVISTA field is
derived from the full COSMOS field \citep[][hereafter W22]{Weaver2022}.
The full COSMOS survey covers the 
wavelength range between 0.2 and 8 $\mu$m, and contains 
1,720,700 sources over 2 degree$^2$.
The COSMOS/UltraVISTA field is within the central 1.5 degree$^2$ of the full COSMOS survey,
which contains about 888,705 objects.
From these 888,705 objects, 405,408 objects are selected as our {\TGS} in the COSMOS/UltraVISTA field
according to the following two criteria:
(1) $0.1 < z \leq 4.0$; (2) S/N of Ks band $\geq 5$.

Next, we collected FIR-submm-radio data from \cite{Jin2018} (hereafter J18).
Similar to the analysis in the GOODS-N,
J18 provided a de-blended FIR ({\it Herschel} 100, 160, 250, 350, and 500 $\mu$m) 
and (sub)millimeter (SCUBA2 850 $\mu$m, AzTEC 1.1 mm, and MAMBO 1.2 mm) photometric catalog
for 191,624 objects in the COSMOS/UltraVISTA field.
This catalog also contains observational data at {\it{Spitzer}} 24 $\mu$m band
and radio detections at 3 GHz from the VLA-COSMOS 3 GHz project with 
an rms noise in the field center of 2.3 $\mu$Jy beam$^{-1}$  \citep{Smolcic2017a}.

Then we cross-matched the FIR-submm-radio catalog (J18) with the 
UV-optical-MIR catalog (W22) by a radius of 0.5 arcsec, 
and 178,494 of 191,624 objects in J18 have UV/optical/MIR counterparts in W22.
The match radius is defined according to the average angular resolution of 3 GHz VLA-COSMOS survey, 
which is 0.75 arcsec \citep{Smolcic2017a}.

Finally, among the above-mentioned 178,494 objects, we selected 7006 objects
as our {\RSS} in the COSMOS/UltraVISTA field according to the following two criteria:
(1) $0.1 < z \leq 4.0$; (2) S/N of 3 GHz radio flux $\geq 5$.

\begin {figure*}
\centering
\begin{adjustbox}{width=1.03\linewidth, right=7in, clip} 
\usetikzlibrary {arrows.meta,bending,positioning}
\begin{tikzpicture}[node distance=18pt]
  \node[draw, fill=green!10]                                            (RAS)               {\begin{tabular}{r} {\RAS} \\ \cdashline{1-1}[1pt/1pt] {\small{\texttt{GOODS-N: 102 sources}}} \\ {\small{\texttt{COSMOS: 881 sources}}} \\ {\small{\texttt{Total: 983 sources}}} \end{tabular}};
  \node[draw, fill=orange!15, rounded corners, left=25pt of RAS]      (ML select)   {\begin{tabular}{c} $F_{\rm 160W} \geq 5 \sigma$ (GOODS) \\ or \\ $F_{\rm Ks} \geq 5 \sigma$ (COSMOS) \end{tabular}};
  \node[draw, fill=blue!10, ellipse, right=25pt of RAS]      (RLF)   {Radio luminosity function$^3$};
  \node[draw, fill=green!5, left=25pt of ML select]                                             (ML source)               {\begin{tabular}{r} {\RAS} \\ used in the ML fitting \\ \cdashline{1-1}[1pt/1pt] {\small{\texttt{GOODS-N:\ \ 98 sources}}} \\ {\small{\texttt{COSMOS: 800 sources}}} \\ {\small{\texttt{Total: 898 sources}}} \end{tabular}};
  \node[draw, fill=green!10, above=565pt of RAS]                                            (TGS)               {\begin{tabular}{r} {\TGS} \\ \cdashline{1-1}[1pt/1pt] {\small{\texttt{GOODS-N:\ \ 29,267 sources}}}  \\ {\small{\texttt{COSMOS: 405,408 sources}}} \\ {\small{\texttt{Total: 434,675 sources}}} \end{tabular}};
  \node[draw, fill=blue!10, ellipse, above=of RAS]               (IRRC)               {IR-radio correlation (IRRC) analysis$^2$};
  \node[draw, fill=green!5, above=of IRRC]         (IRRC source)     {\begin{tabular}{r} Final sample in the IRRC analysis \\ \cdashline{1-1}[1pt/1pt]  {\small{GOODS-N:\ \ \ 509 sources}} \\ {\small{COSMOS: 6985 sources}} \\ \ \ \ \ \ \ {\small{Total: 7494 sources}} \end{tabular}};
  \node[draw, fill=blue!10, ellipse, above=80pt of ML source]                                             (ML)               {Maximum-likelihood (ML) fitting$^4$};
  \node[above=469pt of RAS] (zeropoint) {};
  \node[above=of IRRC source] (node1) {};
  \node[left=90pt of node1] (node2) {};
  \node[right=93pt of node1] (node3) {};
  \node[below=of TGS] (node4) {};
  \node[left=90pt of node4] (node5) {};
  \node[right=93pt of node4] (node6) {};

  \node[draw, fill=gray!10,left=30pt of zeropoint]               (UV GN)               {\begin{tabular}{c} UV-optical-MIR data \\ \cdashline{1-1}[1pt/1pt] {\small{Reference: \cite{Barro2019}}} \\ {\small{Total: 35,445 sources}}  \end{tabular}};
  \node[draw, dotted, thick, left=of UV GN]     (GN)    {\begin{tabular}{c} GOODS-N \\ {\small{(171 arcmin$^2$)}} \end{tabular}};
  \node[draw, fill=gray!10, below=of UV GN]               (Radio GN)               {\begin{tabular}{c} FIR-submm-radio data \\ \cdashline{1-1}[1pt/1pt] {\small{Reference: \cite{Jin2018}}} \\ {\small{(Radio data: 1.4 GHz)}} \\ {\small{Total: 3306 sources}}  \end{tabular}};
  \node[draw, fill=green!5, below=29pt of Radio GN]               (GN UVcounterpart)               {\begin{tabular}{c} 2584 (of 3306) objects with  \\  UV/optical/MIR counterparts \end{tabular}};
  \node[draw, fill=blue!10, ellipse, below=of GN UVcounterpart]      (GN SED)      {\begin{tabular}{c} Broadband SED fitting$^1$ \\ \cdashline{1-1}[1pt/1pt] Best-fit parameters: $L_{\rm TIR}$$^*$, ... \end{tabular}};
  \node[draw, fill=orange!15, rounded corners, below=of GN SED]      (GNradio select)   {$0.1 < z \leq 4$ \& $F_{\rm 1.4\ GHz} \geq 5 \sigma$};
  \node[draw, fill=green!10, below=of GNradio select]                                            (GN RSS)               {\begin{tabular}{c} {\RSS} \\ \cdashline{1-1}[1pt/1pt] {\small{\texttt{509 sources}}} \end{tabular}};
  \node[draw, fill=orange!15, rounded corners, below=of GN RSS]               (GNradio IR select)               {Having $L_{\rm TIR}$ estimates};
  \node[draw, fill=orange!15, rounded corners, above=of UV GN]      (GNUV select)   {$0.1 < z \leq 4$ \& $F_{\rm 160W} \geq 5 \sigma$};

  \node[draw, fill=gray!10, right=30pt of zeropoint]               (UV CS)               {\begin{tabular}{c} UV-optical-MIR data \\ \cdashline{1-1}[1pt/1pt] {\small{Reference: \cite{Weaver2022}}} \\ {\small{Total: 888,705 sources}}  \end{tabular}};
  \node[draw, dotted, thick, right=of UV CS]     (CS)    {\begin{tabular}{c} COSMOS \\ {\small{(UltraVISTA;}} \\ {\small{1.5 degree$^2$)}} \end{tabular}};
  \node[draw, fill=gray!10, below=of UV CS]               (Radio CS)               {\begin{tabular}{c} FIR-submm-radio data \\ \cdashline{1-1}[1pt/1pt] {\small{Reference: \cite{Jin2018}}} \\ {\small{(Radio data: 3 GHz)}} \\ {\small{Total: 191,624 sources}} \\ {\small{SED fitting: $L_{\rm TIR}$$^*$ estimates}}  \end{tabular}};
  \node[draw, fill=green!5, below=of Radio CS]               (CS UVcounterpart)               {\begin{tabular}{c} 178,494 (of 191,624) objects with  \\  UV/optical/MIR counterparts \end{tabular}};
  \node[draw, fill=orange!15, rounded corners, below=78pt of CS UVcounterpart]      (CSradio select)   {$0.1 < z \leq 4$ \& $F_{\rm 3\ GHz} \geq 5 \sigma$};
  \node[draw, fill=green!10, below=of CSradio select]                                            (CS RSS)               {\begin{tabular}{c} {\RSS} \\ \cdashline{1-1}[1pt/1pt] {\small{\texttt{7006 sources}}} \end{tabular}};
  \node[draw, fill=orange!15, rounded corners, below=of CS RSS]               (CS parameter)               {Having $L_{\rm TIR}$ estimates};
  \node[draw, fill=orange!15, rounded corners, above=of UV CS]      (CSUV select)   {$0.1 < z \leq 4$ \& $F_{\rm Ks} \geq 5 \sigma$};

\draw[arrows = {-Stealth[scale=1.5]}] (TGS) -- (TGS-|ML) -> (ML);
\draw[arrows = {-Stealth[scale=1.5]}] (UV GN) -- (GNUV select);
\draw[arrows = {-Stealth[scale=1.5]}] (GN) -- (UV GN);
\draw[arrows = {-Stealth[scale=1.5]}] (UV GN) -- node[right] {Cross-match} (Radio GN);
\draw[arrows = {-Stealth[scale=1.5]}] (Radio GN) -- node[right] {Yes} (GN UVcounterpart);
\draw[arrows = {-Stealth[scale=1.5]}] (GN UVcounterpart) -- (GN SED);
\draw[arrows = {-Stealth[scale=1.5]}] (GN SED) -- (GNradio select);
\draw[arrows = {-Stealth[scale=1.5]}] (GNradio select) -- node[right] {Yes} (GN RSS);
\draw[arrows = {-Stealth[scale=1.5]}] (GN RSS) -- (GNradio IR select);

\draw[arrows = {-Stealth[scale=1.5]}] (IRRC source) -- (IRRC);
\draw[arrows = {-Stealth[scale=1.5]}] (IRRC) -- (RAS);
\draw[] (GNradio IR select) -- (node2.center) -- node[above] {Yes} (node1.center) -- node[above] {Yes} (node3.center) -- (CS parameter);
\draw[arrows = {-Stealth[scale=1.5]}] (node1.center) -- (IRRC source);
\draw[] (GNUV select) -- (node5.center) -- node[above] {Yes} (node4.center) -- node[above] {Yes} (node6.center) -- (CSUV select);
\draw[arrows = {-Stealth[scale=1.5]}] (node4.center) -- (TGS);

\draw[arrows = {-Stealth[scale=1.5]}] (UV CS) -- (CSUV select);
\draw[arrows = {-Stealth[scale=1.5]}] (CS) -- (UV CS);
\draw[arrows = {-Stealth[scale=1.5]}] (UV CS) -- node[right] {Cross-match} (Radio CS);
\draw[arrows = {-Stealth[scale=1.5]}] (Radio CS) -- node[right] {Yes} (CS UVcounterpart);
\draw[arrows = {-Stealth[scale=1.5]}] (CS UVcounterpart) -- (CSradio select);
\draw[arrows = {-Stealth[scale=1.5]}] (CSradio select) -- node[right] {Yes} (CS RSS);
\draw[arrows = {-Stealth[scale=1.5]}] (CS RSS) -- (CS parameter);
\draw[arrows = {-Stealth[scale=1.5]}] (RAS) -- (RLF);
\draw[arrows = {-Stealth[scale=1.5]}] (RAS) -- (ML select);
\draw[arrows = {-Stealth[scale=1.5]}] (ML select) -- node[above] {Yes} (ML source);
\draw[arrows = {-Stealth[scale=1.5]}] (ML source) -- (ML);

\end{tikzpicture}
\end{adjustbox}
\caption{Flowchart of the matching processes and sample selections (see details in Section \ref{sec:data}).
$^1$ See details about broadband SED fitting in Section \ref{sec:LIRM} and Appendix \ref{sec:SED}.
$^2$ See details about IR-radio correlation (IRRC) analysis in Section \ref{sec:REAGNpaper}.
$^3$ See details about radio luminosity function in Section \ref{sec:RLF}.
$^4$ See details about Maximum-likelihood (ML) fitting method used for calculating the probability of hosting a radio-excess AGN in Section \ref{sec:Fagntotalsummary}. $^*$$L_{\rm TIR}$: rest-frame total IR luminosity between 8 and 1000 $\mu$m.
 \label{fig:flowchart}}
\end{figure*}
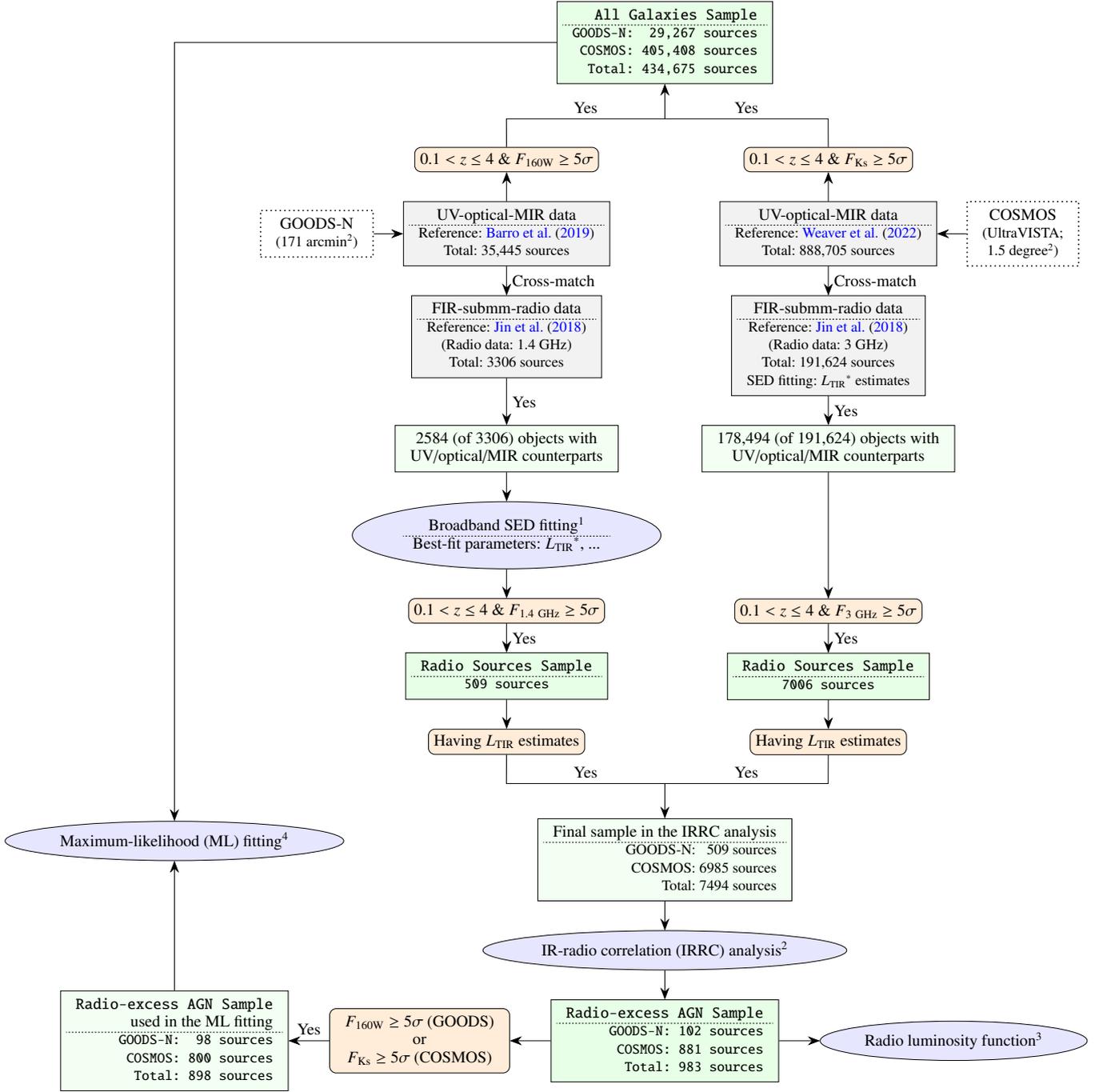

\section{Galaxy property estimation}
\label{sec:parameter}

\subsection{Rest-frame 8-1000 $\mu$m IR luminosity}
\label{sec:LIRM}
{\it GOODS-N field:} For 2584 objects in the GOODS-N field
which have multi-wavelength data from UV to FIR/sub-mm to radio bands,
we estimated the total IR luminosity from 8 to 1000 $\mu$m
through spectral energy distribution (SED) fitting with 
Code Investigating GALaxy Emission 
\citep[\textsc{cigale} 2022.0;][]{Burgarella2005,Noll2009,Boquien2019,Yang2020,Yang2022}.
We refer readers to Appendix \ref{sec:SED} for more details about the SED fitting.

{\it COSMOS/UltraVISTA field:} For sources in the COSMOS/UltraVISTA field,
we use 8--1000 $\mu$m luminosity estimated by the SED fitting in J18.
Their SED components consist of a stellar component from \cite{Bruzual2003} with a Small
Magellanic Cloud attenuation law, dust continuum emission based on \cite{Draine2007}, a MIR AGN torus component
form \cite{Mullaney2011}, and a power-law radio component.
In addition, our work used the same \cite{Chabrier2003} IMF as J18.

All the objects of the {\RSS} in the GOODS-N field have available IR luminosity estimates.
6985 (of 7006) objects of the {\RSS} in the COSMOS/UltraVISTA field have $L_{\rm TIR}$ estimates from J18 ($\sim 99.7\%$). 
Moreover, we have verified that different methods
give similar IR luminosity measurements. For example, at $0.1 < z < 1$, averaged IR luminosities of galaxies with
$10^{10.5} < M_\star < 10^{11.5}\ M_\odot$ in the GOODS-N and COSMOS/UltraVISTA fields are 
$3.2\times 10^{37}$ and $2.6\times 10^{37}$ W, repectively.

\begin{figure}[t!]
\includegraphics[width=\linewidth, clip]{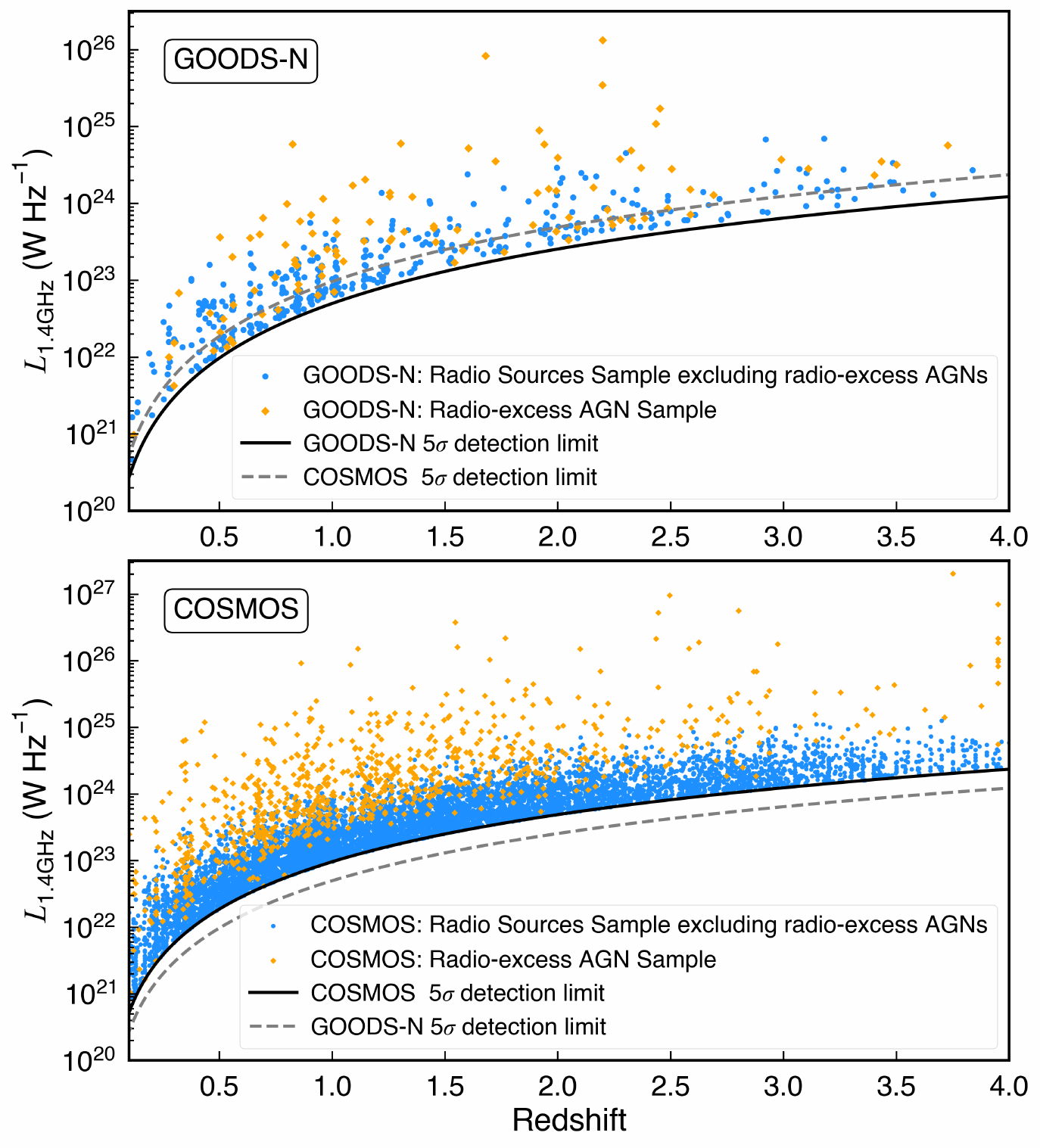}
\caption{Distribution of rest-frame 1.4 GHz luminosity 
of {\RSS} (blue circles) and {\RAS} (orange diamonds) with redshift
for the GOODS-N ({\it top}) and COSMOS/UltraVISTA ({\it bottom}) fields.
The $5\sigma$ detection limit for the COSMOS/UltraVISTA field
with $\sigma = 2.3\ \mu {\rm Jy}\ {\rm beam}^{-1}$ at 3 GHz \citep{Smolcic2017a}
is shown as a gray dashed line in the top panel and as a black solid line in the bottom panel.
The $5\sigma$ detection limit for the GOODS-N field with $\sigma = 2.2\ \mu {\rm Jy}$ at 1.4 GHz \citep{Owen2018}
is shown as a black solid line in the top panel and as a gray dashed line in the bottom panel.
All the detection limits are calculated by Equation \ref{equ:LRrest} assuming a fixed spectral index of $-0.8$ 
(see details in Section \ref{sec:L1.4GHz}).
\label{fig:radiolimit}}
\end{figure}

\subsection{Rest-frame 1.4 GHz radio luminosity}
\label{sec:L1.4GHz}
Throughout this work, the radio spectrum of each radio source
is assumed to follow a simple power law shape of $S_{\nu} \propto \nu^{\alpha}$,
where $S_\nu$ is the flux density at frequency $\nu$ and $\alpha$ is the spectral index
which is assumed to be $-0.8$ in this work.
This simple power law assumed for radio spectrum is a widely used approximation
due to insufficient radio data,
and it is worth noting that adopting a single value of $\alpha$ may bring
a scatter or bias to the calculation of radio luminosity \citep[e.g.,][]{Gim2019}.
The rest-frame 1.4 GHz luminosity (in the unit of $\rm{W}\ \rm{Hz}^{-1}$) of the {\RSS} can be calculated by 
\begin{equation}
\label{equ:LRrest}
L_{1.4 \rm{GHz}, \rm{rest}} = \frac{4\pi D_L^2}{(1+z)^{1+\alpha}} \left(\frac{1.4\ \rm{GHz}}{\nu_{\rm obs}}\right)^{\alpha} S_{\nu, \rm{obs}},
\end{equation}
where $D_L$ is the luminosity distance (in the unit of meter), $z$ is the redshift, 
$\nu_{\rm obs}$ is the observed frequency (in the unit of GHz),
and $S_{\nu, \rm{obs}}$ is the observed integrated flux densities at the observed frequency $\nu_{\rm obs}$ \citep{Ceraj2018}.
For the GOODS-N field, we use the observed 1.4 GHz flux to calculate the rest-frame 1.4 GHz luminosity,
while for the COSMOS/UltraVISTA field, we use the observed flux at 3 GHz.
We show the distribution of rest-frame 1.4 GHz luminosity of 
our {\RSS} with redshift in Fig. \ref{fig:radiolimit}.

\subsection{Stellar mass and UVJ magnitude}
For the {\TGS} in the GOODS-N field, 
their stellar masses and rest-frame UVJ magnitudes are derived from \cite{Barro2019} where stellar masses 
are estimated by SED fitting with the codes \texttt{FAST} \citep{Kriek2009} 
and \texttt{Synthesizer} \citep{PerezGonzalez2005,PerezGonzalez2008},
and rest-frame UVJ luminosities are estimated by \texttt{EAZY} \citep{Brammer2008}.
For the {\TGS} in the COSMOS/UltraVISTA field,
their stellar masses and rest-frame UVJ magnitudes are derived from \cite{Weaver2022} where stellar masses are 
estimated by \texttt{LePhare} \citep{Arnouts2002,Ilbert2006}
and rest-frame UVJ luminosities are estimated by \texttt{EAZY} \citep{Brammer2008}.
Our work and these aforementioned works used the same \cite{Chabrier2003} IMF.
In addition, to examine whether the above different methods will bring measurement bias for stellar mass,
we also compare stellar masses given by the above methods 
($M_{\star}^{\rm \texttt{FAST}/\texttt{LePhare}}$) with those obtained by the broadband SED fitting
($M_{\star}^{\rm SED}$; see details in Appendix \ref{sec:SED} for the GOODS-N field and see details in \citealt{Jin2018}
for the COSMOS field).
The logarithm of the ratio between $M_{\star}^{\rm \texttt{FAST}/\texttt{LePhare}}$ and $M_{\star}^{\rm SED}$
($\log [M_{\star}^{\rm \texttt{FAST}/\texttt{LePhare}}/M_{\star}^{\rm SED}]$)
is around $0 \pm 0.3$ dex.
This result indicates that different methods
give a consistent measurements for stellar mass.
Given that only the {\RSS} and {\RAS} have available broadband SED fitting results while the {\TGS} do not have these measures,
we use the stellar mass measured by \texttt{FAST} or \texttt{LePhare} in this work.

\subsection{Galaxy type: SFGs and QGs}
\label{sec:UVJ}
Color-magnitude or color-color criteria are an effective method to select SFGs and QGs,
which does not require accurate measurements for star formation rate and stellar mass \citep[e.g.,][]{Williams2009}.
In this work, we use UVJ selection criteria in \cite{Schreiber2015}
to separate all our samples into SFGs and QGs at all redshift and stellar masses:
\begin{equation}
\rm{quiescent} = 
  \begin{cases}
    U-V > 1.3\ \rm{and} \\
    V-J < 1.6\ \rm{and} \\
    U-V > 0.88\times (V-J)+0.49 \\
  \end{cases}
\end{equation}

\begin{equation}
\operatorname{star-forming}\ =  
  \begin{cases}
    U-V \leq 1.3\ \rm{or} \\
    V-J \geq 1.6\ \rm{or} \\
    U-V \leq 0.88\times (V-J)+0.49 \\
  \end{cases}
\end{equation}

\begin{figure}
\includegraphics[width=\linewidth, clip]{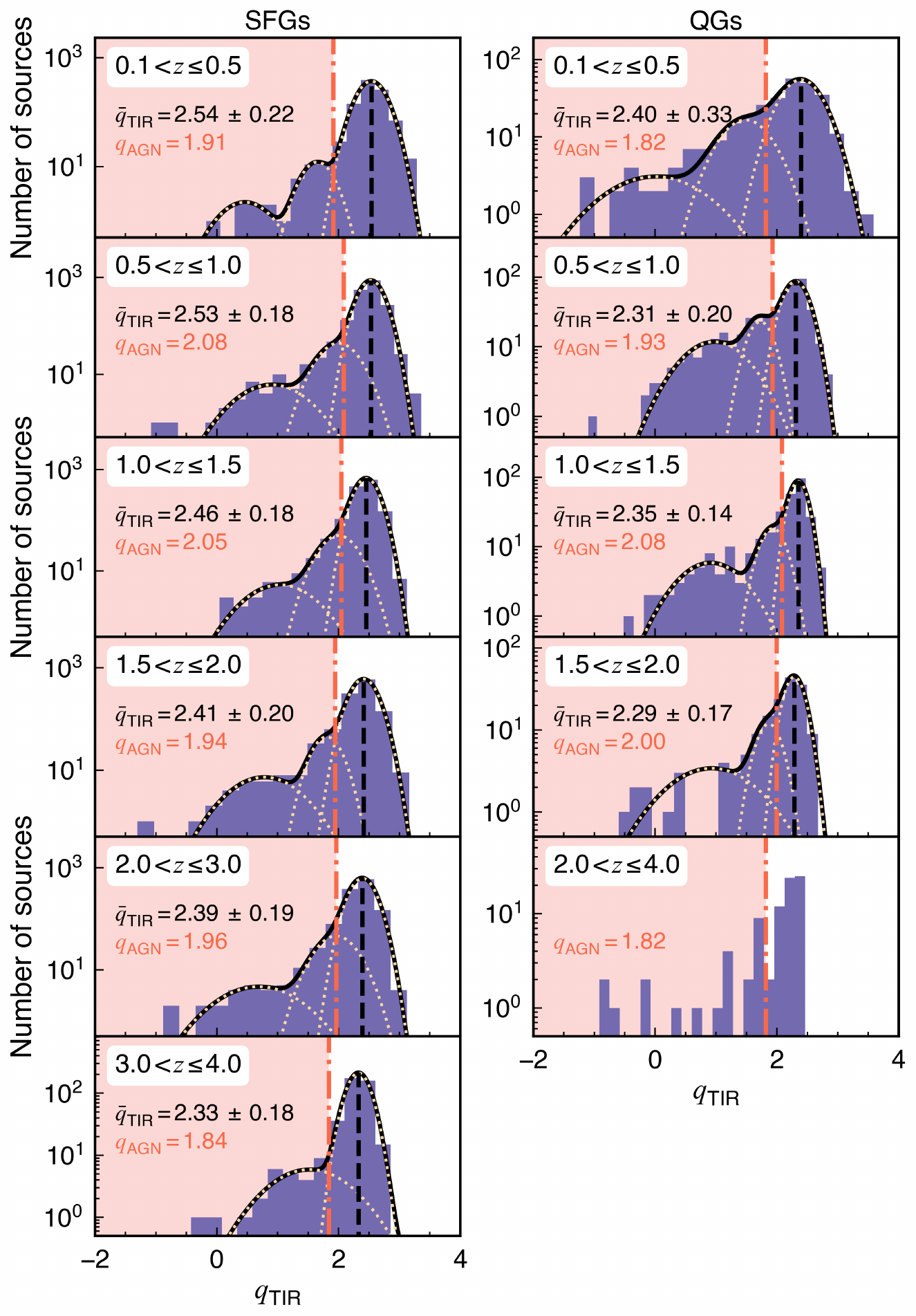}
\caption{Distribution of IR-radio-ratio ($q_{\rm TIR}$) in SFGs ({\it left} column)
and QGs ({\it right} column) at different redshift bins 
(from {\it top} to {\it bottom} panels).
The $q_{\rm TIR}$ distributions are obtained by merging 
{\RSS} from the GOODS-N and COSMOS/UltraVISTA fields together.
Each yellow dotted curve represents each single gaussian model,
while the black solid curve represents the best-fit model to the entire $q_{\rm TIR}$ distribution.
The vertical black dashed line (corresponds to the $\bar{q}_{\rm TIR}$ value in the figure)
represents the peak position of the highest gaussian component.
The vertical red dash-dotted line (corresponds to the $q_{\rm AGN}$ value in the figure) 
represents the cross point between the highest gaussian component and the 
second highest gaussian component, which is used as the threshold 
to separate SFGs and radio-excess AGNs in this work.
For QGs at $2.0 < z \leq 4.0$,
it is difficult to find a best-fit model due to the low source number, 
so here we use the threshold at $0.1 < z \leq 0.5$,
which does not have a significant effect to the results.
\label{fig:qIRdis}}
\end{figure}

\section{Radio-excess AGN}
\label{sec:REAGNpaper}

The excess radio emission from AGNs will make these types of sources deviate from
the IRRC of SFGs.
Therefore, AGN activities can be identified by these excess radio emission.
Following \cite{Helou1985}, we define the logarithmic ratio between total IR emission (8--1000 $\mu$m; TIR)
 and radio emission as
\begin{equation}
q_{\rm{TIR}} = \log \left( \frac{L_{\rm{TIR}}}{L_{1.4 \rm{GHz}, \rm{rest}} \times 3.75 \times 10^{12}\ \rm{Hz}} \right),
\end{equation}
where $L_{\rm{TIR}}$ is the total IR luminosity and 
$3.75 \times 10^{12}\ \rm{Hz}$ is the frequency at the center of FIR band ($\lambda = 80 \mu \rm{m}$).
In this work, we consider a broad redshift range from 0.1 to 4,
which are divided into 6 redshift bins:
$0.1 < z \leq 0.5$, $0.5 < z \leq 1.0$, $1.0 < z \leq 1.5$,
$1.5 < z \leq 2.0$, $2.0 < z \leq 3.0$, and $3.0 < z \leq 4.0$.
In this section, we focus on the {\RSS}
and then use $q_{\rm{TIR}}$ distribution to select 
AGNs with excess radio emission which are called as radio-excess AGNs hereafter.

\subsection{$q_{\rm{TIR}}$ distribution and radio-excess AGN selection}
\label{sec:qTIRagn}
In this analysis we merge the radio sources in the GOODS-N 
and COSMOS/UltraVISTA fields together according to the following reasons.
(1) Source numbers in each redshift bin for the GOODS-N field are small,
which may result in a large uncertainty in determining the AGN selection threshold.
(2) In each redshift bin, $q_{\rm{TIR}}$ distributions in the GOODS-N and COSMOS fields
have similar $q_{\rm{TIR}}$ distributions and $\bar{q}_{\rm{TIR}}$ values
(see Fig. \ref{fig:qIRfield}).
The $q_{\rm{TIR}}$ distribution consists of a gaussian component peaked at $\bar{q}_{\rm{TIR}}$ and 
an extended tail towards lower $q_{\rm{TIR}}$,
which is consistent with previous works \citep[e.g.,][]{DelMoro2013}.
The gaussian component is attributed to the IRRC of SFGs,
while the extended tail is thought to be associated with the extra radio emission from AGN.
Therefore, to ensure the same selection criteria and reduce the uncertainty of the AGN selection threshold,
we combine the data from the two fields together.
The consistent radio luminosity functions among these two fields also indicate that
our selection method is plausible (see details in Section \ref{sec:RLF}).
Next we analyze the $q_{\rm{TIR}}$ distribution in SFGs and QGs, respectively.

The entire $q_{\rm{TIR}}$ distribution can be described by multiple gaussian models 
(see the yellow dotted curves in Fig. \ref{fig:qIRdis}).
We used $f$-test probability at $\geq$ 95\% confidence level to decide how many models are required.
The final best-fit model are shown in Fig. \ref{fig:qIRdis} with black solid curves.
As we mentioned before, the highest gaussian peak and the extended low-$q_{\rm{TIR}}$ tail
are thought to be contributed by SFGs and AGNs, respectively.
Therefore, here we define the cross point between the highest gaussian 
and the second highest gaussian components as the threshold to separate the SFGs and radio-excess AGNs 
(see the $q_{\rm AGN}$ values and red dash-dotted lines in Fig. \ref{fig:qIRdis}).
It means that a radio source will be selected as a radio-excess AGN 
if its $q_{\rm{TIR}}$ value is lower than this cross point ($q_{\rm AGN}$).
It is worth noting that the $q_{\rm{TIR}}$ distributions of QGs also show a highest peak similar to SFGs.
However, QGs are not expected to  
have this peak due to little or no star formation.
One possible reason may be that UVJ color selection might classify part of 
massive SFGs with relatively lower star formation rate (SFR) as QGs \citep{Popesso2023}.
We also found that nearly half of UVJ-selected QGs (without AGN) in the {\RSS} locate very close
to the dividing line in the UVJ diagram, which indicates that
UVJ selection might not accurately classify these sources.
In addition, it is worth noting that QGs in the {\TGS} do not show such a distribution in the UVJ diagram.
Therefore, another possible reason may be that radio detection favors sources with higher star formation, even among QGs.
Radio-excess AGNs hosted by QGs do not show a prominent populations close to the dividing line in the UVJ diagram,
which may be due to that radio-excess AGNs are selected by a threshold ($q_{\rm{AGN}}$) that is significantly lower than the mean $q_{\rm{TIR}}$ of SFGs.
This may explain why QGs show a highest $q_{\rm{TIR}}$ peak in Fig. \ref{fig:qIRdis}.
In addition,
the difference between the second highest and the third highest gaussian component in the $q_{\rm TIR}$ distribution might be
associated with different types of radio AGNs, which is beyond the scope of this work and will be studied in our future works.

Finally, we selected 983 radio-excess AGNs at $0.1 < z \leq 4$ as our {\RAS}
which contains 102 sources in the GOODS-N field
and 881 sources in the COSMOS/UltraVISTA field 
(see Fig. \ref{fig:radiolimit}).

\begin{figure}
\includegraphics[width=\linewidth, clip]{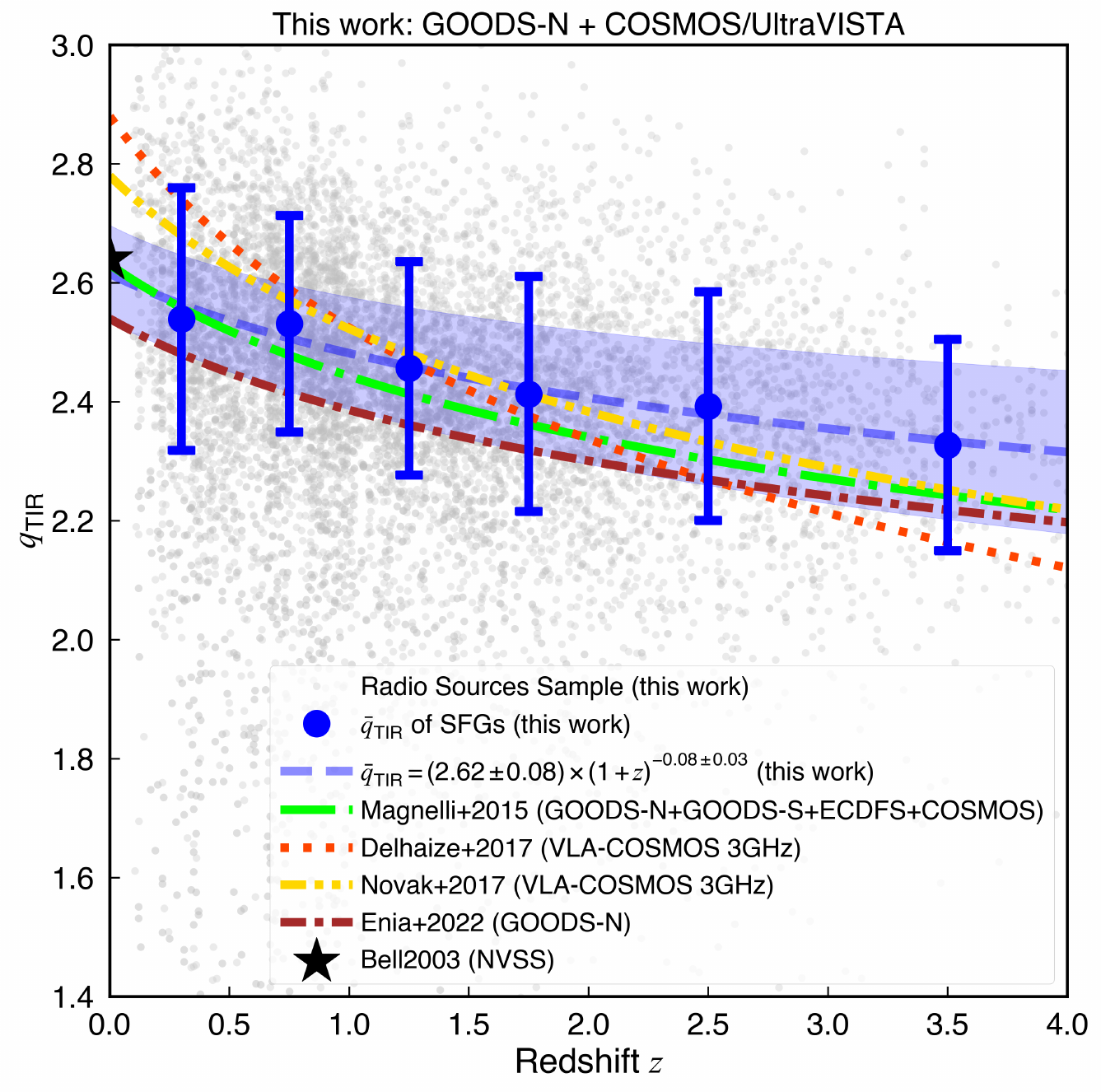}
\caption{The distribution of $q_{\rm TIR}$ with redshift.
The gray points show the {\RSS} in the GOODS-N and COSMOS/UltraVISTA fields
(see details in Section \ref{sec:data}).
The blue points represent the $\bar{q}_{\rm TIR}$ values of SFGs
which are shown in Fig. \ref{fig:qIRdis} and in Section \ref{sec:qTIRagn}.
The blue dashed line represents the best-fit model to the data of SFGs,
while the shaded blue region is the associated 1$\sigma$ uncertainty of the best-fit model.
The green dash-dotted line represents the best-fit model from \cite{Magnelli2015} 
using data in the GOODS-N, GOODS-S, ECDFS, and COSMOS fields. 
Both the red dotted line \citep{Delhaize2017} and yellow dash-dotted line \citep{Novak2017} 
represent the best-fit model obtained with the VLA-COSMOS 3 GHz project but using different
source selection criteria. 
The dark-red dash-dotted line represents the best-fit model from \cite{Enia2022}
based on data from the GOODS-N field.
The black star represents the $q_{\rm TIR}$ value for SFGs in the local universe from \cite{Bell2003}.
\label{fig:qIRevo}}
\end{figure}

\subsection{$q_{\rm{TIR}}$ evolution}
\label{sec:qIRevolution}
As Fig. \ref{fig:qIRevo} shows, 
$\bar{q}_{\rm{TIR}}$ presents a weak evolution with redshift,
which is in the form of
\begin{equation}
\bar{q}_{\rm{TIR}} = (2.62\pm 0.08)\times (1+z)^{-0.08\pm 0.03}.
\end{equation}
Our result is almost consistent with previous works \citep[see Fig. \ref{fig:qIRevo};][]{Bell2003,Magnelli2015,Delhaize2017,Novak2017,Enia2022}.
\cite{Bell2003} obtained $q_{\rm TIR}$ value for SFGs in the local universe.
\cite{Magnelli2015} combined data from multiple fields of the GOODS-N, GOODS-S, ECDFS, and COSMOS up to $z \sim 2$
with rms noises larger than 3.9 $\mu {\rm Jy}$ at 1.4 GHz.
\cite{Delhaize2017} and \cite{Novak2017} both used the VLA-COSMOS 3 GHz project but used different source selection criteria.
\cite{Enia2022} focused on SFGs in the GOODS-N field.
Compared to their works, we use simultaneously deeper and larger radio surveys, 
and more precise IR luminosity based on de-blended IR photometry.
The slight difference between our results and previous works might be due to the different selection method for sources.
In addition, the averaged standard deviation (1$\sigma$ error of $q_{\rm TIR}$ in Fig. \ref{fig:qIRevo}) 
of $q_{\rm TIR}$ distribution for SFGs is around 0.18 
which is consistent with the previous works \citep[e.g.,][]{DelMoro2013}.

More recently, \cite{Delvecchio2021} found that the IRRC of SFGs strongly depends on stellar mass ($M_\star$),
while \cite{An2021} found a weak dependence.
To examine whether considering $M_\star$-dependence or not affects our results,
we make some simple tests.
For the {\RSS} in the COSMOS/UltraVISTA field,
at $0.1 < z < 1$, $\bar{q}_{\rm TIR}$ is around 2.55 and 2.53 
for low-mass($10^{10} < M_\star < 10^{10.5}\ M_\odot$) and
high-mass ($10^{10.5} < M_\star < 10^{11}\ M_\odot$) SFGs, respectively.
For the {\RSS} in the GOODS-N field,
at $0.1 < z < 1$, $\bar{q}_{\rm TIR}$ is around 2.56 and 2.51 
for low-mass ($10^{10} < M_\star < 10^{10.5}\ M_\odot$) and
high-mass ($10^{10.5} < M_\star < 10^{11}\ M_\odot$) SFGs, respectively.
These results indicate that for our samples, 
$\bar{q}_{\rm TIR}$ of SFGs shows no significant dependence on stellar mass within the uncertainties of $q_{\rm TIR}$.
Therefore, disregarding the $M_*$-dependent IRRC may have no significant effect on our results.

\begin{figure*}
\includegraphics[width=\linewidth, clip]{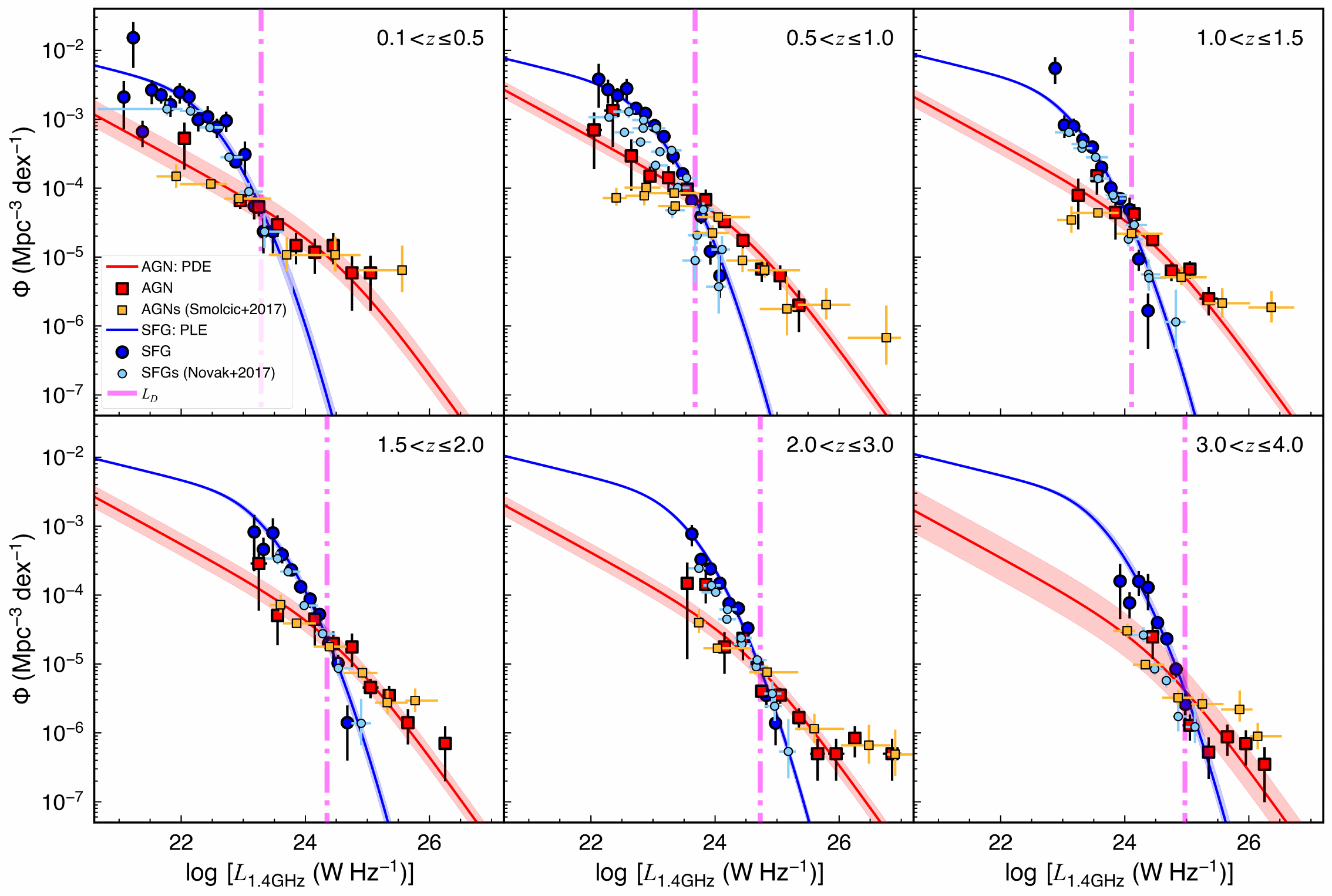}
\caption{The 1.4 GHz radio luminosity function of SFGs (blue circles) 
and radio-excess AGNs (red squares) in different redshift bins.
For SFGs, blue solid line represents the best-fit PLE model 
which is the best model to describe the data in this work.
For radio-excess AGNs, red solid line represents the best-fit PDE model 
which is the best model to describe the data. 
The magenta vertical dash-dotted line represents the turnover luminosity between the 
radio luminosity function of SFGs and radio-excess AGNs (see details in Section \ref{sec:turnoverL}).
The light-sky-blue circles show the results for SFGs from the VLA-COSMOS 3 GHz project \citep{Novak2017} while the gold squares show the
results for radio AGNs from the VLA-COSMOS 3 GHz project \citep{Smolcic2017b}.
It is worth noting that we use different definition for redshift bins comparing to both \cite{Novak2017} and \cite{Smolcic2017b}
especially for the highest redshift bin.
\label{fig:LFtotal}}
\end{figure*}

\section{Radio luminosity function and their cosmic evolution}
\label{sec:RLF}

In this section, we will construct radio luminosity function (RLF) for SFGs and radio-excess AGNs 
which are separated in Section \ref{sec:REAGNpaper}.
Firstly, we estimate the combined 1.4 GHz RLF
from the GOODS-N and COSMOS/UltraVISTA fields in Section \ref{sec:RLFobs}.
We then show the fitting procedures and results for the evolution of SFG RLFs
(see Section \ref{sec:RLFSFG}) and AGN RLFs (see Section \ref{sec:RLFAGN}), respectively.
In this work we do not consider the star-formation contamination for the 1.4 GHz radio luminosity of AGN.
We have verified that for most of our {\RAS}, 
radio luminosity from the star formation is not significant compared to luminosity from AGN.

\subsection{Construct RLF out to $z \sim 4$}
\label{sec:RLFobs}
Using the 1/$V_{\rm max}$ method \citep{Schmidt1968}, 
RLF in each 1.4 GHz luminosity bin and each redshift bin is calculated by 
\begin{equation}
\Phi (L,z) = \frac{1}{\Delta \log L} \sum_i \frac{1}{\frac{\Omega}{4 \pi} \times V_{\rm max,i} \times w_i},
\label{equ:RLFfunction}
\end{equation}
where $\Delta \log L$ is the size of the 1.4 GHz luminosity bin, 
$\Omega$ is the observed area (171 arcmin$^2$ for the GOODS-N field,
and 1.5 degree$^2$ for the COSMOS/UltraVISTA field),
$V_{\rm max,i}$ is the co-moving volume of the $i$th source that 
is defined as $V_{\rm max,i} = V_{z_{\rm max,i}} - V_{z_{\rm min}}$,
and $w_i$ is the completeness and bias correction factor of the $i$th source.
Further, $V_{z_{\rm max,i}}$ is the co-moving volume at the maximum redshift where the $i$th source
can be observed given the 1.4 GHz flux detection limit 
(the maximum value of $z_{\rm max}$ is equal to the upper limit of each redshift bin) 
and $V_{z_{\rm min}}$ is the co-moving volume
at the lower boundary of each redshift bin.
The parameter $w_i$ is the flux density completeness of our catalogs 
which takes into account the effects of sensitivity limit.
The completeness and bias corrections of the COSMOS and GOODS-N fields are 
derived from \cite{Smolcic2017a} and \cite{Enia2022}, respectively, which are
estimated by Monte Carlo simulations where mock sources 
are inserted in and retrieved from the mosaic.
The uncertainty of RLF in each luminosity bin and each redshift bin can be defined as
\begin{equation}
\sigma_{\Phi} (L,z) = \frac{1}{\Delta \log L} \sqrt{\sum_i \left(\frac{1}{\frac{\Omega}{4 \pi} \times V_{\rm max,i} \times w_i}\right)^2},
\label{equ:RLFerror}
\end{equation}

We used Equations \ref{equ:RLFfunction} and \ref{equ:RLFerror} to 
calculate RLFs and their uncertainties for the sources in the GOODS-N 
and COSMOS/UltraVISTA, respectively (see Fig. \ref{fig:RLFfield}).
For both SFGs and radio-excess AGNs, RLFs in these fields present generally consistent results.
At the faintest end of the lowest redshift bin ($0.1 < z \leq 0.5$), SFG RLF of the COSMOS/UltraVISTA field presents a slight decline comparing to that of the GOODS-N field (see Figure \ref{fig:RLFfield}).
This may be due to that the higher spatial resolution of the radio survey in the COSMOS/UltraVISTA field (see details in Section \ref{sec:data}) results in more flux losses for the extended sources especially for the faint populations at low redshift.
In order to well constrain the RLFs,
in the following sections we use the averaged SFG RLFs in the GOODS-N and COSMOS/UltraVISTA fields,
and use the averaged AGN RLFs in the two fields to make further analysis (see Fig. \ref{fig:LFtotal}).
In addition, we also correct the classification purity for the 
radio-excess AGN in these three fields (see details in Section \ref{sec:REAGNpurity}).

\begin{figure*}[t!]
\centering
\includegraphics[width=0.85\linewidth, clip]{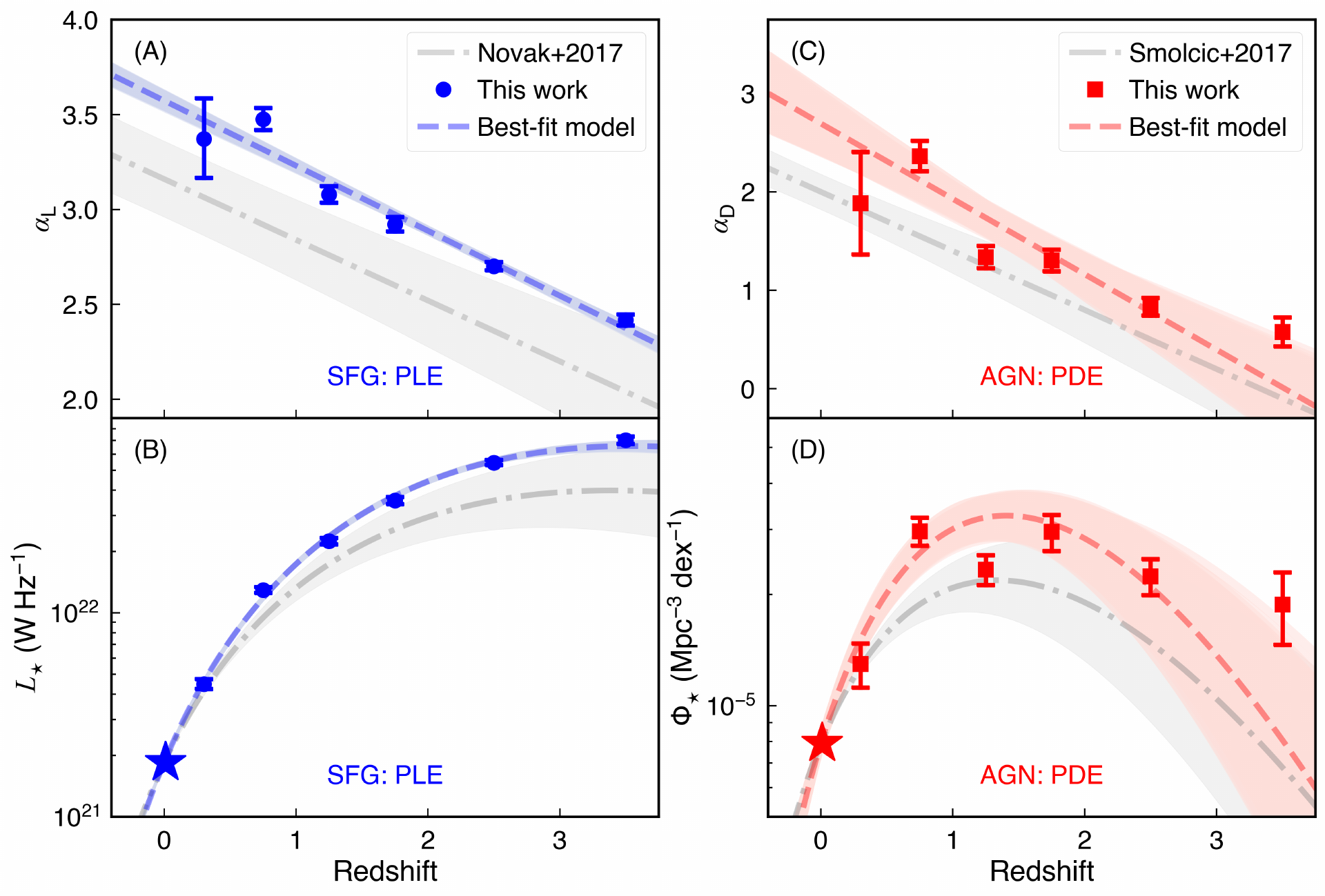} \\
\includegraphics[width=0.85\linewidth, clip]{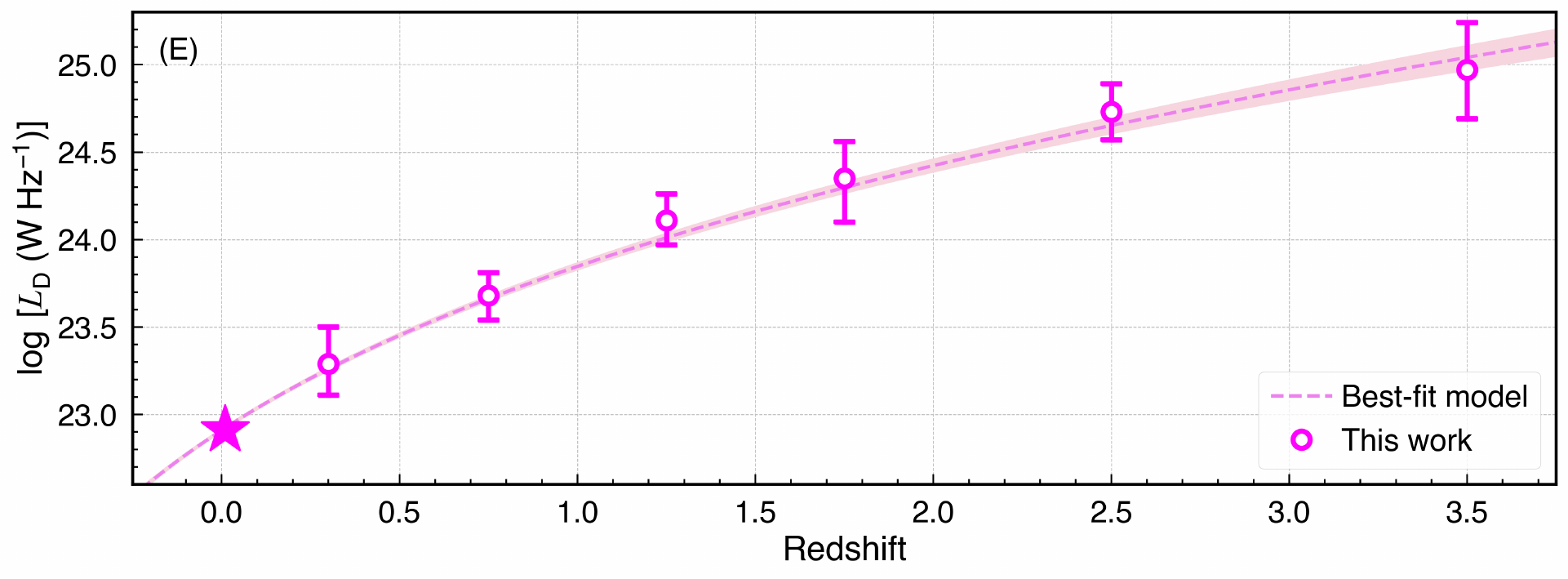}
\caption{Evolution of the best-fit parameters for the 1.4 GHz radio radio luminosity function (RLF) with redshift:
evolution parameter ($\alpha_{\rm L}$) of the pure luminosity evolution (PLE) for SFGs ({\it Panel A}),
turnover position ($L_\star$) of the PLE for SFGs which is calculated by Equation \ref{equ:Lstar} ({\it Panel B}),
evolution parameter ($\alpha_{\rm D}$) of the pure density evolution (PDE) for radio-excess AGN ({\it Panel C}),
turnover normalization ($\Phi_{\star}$) of the PDE for radio-excess AGN
which is calculated by Equation \ref{equ:Phistar} ({\it Panel D}),
turnover luminosity ($L_{\rm D}$) above which the 
number density of radio-AGNs surpasses that of SFGs ($L_{\rm D}$) ({\it Panel E}).
The gray dash-dotted lines in Panels A and B represent the results for SFGs from \cite{Novak2017},
while the gray regions represent the corresponding 1-$\sigma$ uncertainty.
The gray dash-dotted lines in Panels C and D represent the results for radio-excess AGNs from \cite{Smolcic2017b},
while the gray regions represent the corresponding 1-$\sigma$ uncertainty.
The stars in Panels B and D represent the local values obtained by \cite{Novak2017} and \cite{Mauch2007}, respectively.
The magenta star in Panel E represents the local $L_{\rm D}$ 
obtained by the cross point between the local SFGs RLF from \cite{Novak2017} 
and the local AGN RLF from \cite{Mauch2007}.
The best-fit values for $\alpha_{\rm L}$, $\alpha_{\rm D}$, $L_\star$, $\Phi_\star$, and $L_{\rm D}$
are summarized in Table \ref{tab:RLFevolution}.
\label{fig:LFevo}}
\end{figure*}

\subsection{Fit the evolution of the SFG RLF}
\label{sec:RLFSFG}
We assume a modified-Schechter function \citep{Saunders1990,Smolcic2009,Gruppioni2013,Novak2017} 
to describe the SFG RLF.
The local RLF of SFGs \citep{Novak2017} is defined as
\begin{equation}
\label{eqa:phi0}
\Phi_{0}^{\rm SFG} (L) 
= \Phi_{\star,0}\left(\frac{L}{L_{\star,0}}\right)^{1-\alpha}\exp\left[-\frac{1}{2\sigma^2}\log^2\left(1+\frac{L}{L_{\star,0}}\right)\right],
\end{equation}
where $L_{\star,0}= 1.85\times 10^{21}\ \rm{W}\ \rm{Hz}^{-1}$ is the turnover position of the local SFG RLF, 
$\Phi_{\star,0}= 3.55\times 10^{-3}\ \rm{Mpc}^{-3}\ \rm{dex}^{-1}$ is the local turnover normalization, 
and $\alpha=1.22$ and $\sigma= 0.63$ are used to fit the distribution in the faint and bright ends, respectively.

SFG RLFs have been shown to evolve with redshift \citep[e.g.,][]{Novak2017,Ocran2020,Cochrane2023}.
Given that it is difficult to simultaneously constrain the evolutions of all the parameters,
we fix the $\alpha$ and $\sigma$ at all cosmic times to those of the local RLF \citep{Novak2017} in the fit.
It means that we assume an unchanged RLF shape at all cosmic times, and 
only allow the turnover position ($L_{\star}$) and turnover
normalization ($\Phi_{\star}$) to change with redshift.
In reality, $\alpha$ and $\sigma$ might change with redshift.

Both $L_{\star}$ evolution and $\Phi_{\star}$ evolution can be described by 
a simple power law:
\begin{equation}
\label{equ:Lstar}
L_{\star} = (1+z)^{\alpha_{L}} \times L_{\star,0}\ 
\end{equation}
and
\begin{equation}
\label{equ:Phistar}
\Phi_{\star} = (1+z)^{\alpha_{D}} \times \Phi_{\star,0},
\end{equation}
respectively.
Therefore, the redshift-evolved SFG RLF (luminosity and density evolution; hereafter ``LDE'') is defined as
\begin{equation}
\Phi^{\rm SFG}_{\rm LDE}(L, z, \alpha_{L}, \alpha_{D})
= (1+z)^{\alpha_{D}}\times \Phi_{0}^{\rm SFG}\left(\frac{L}{(1+z)^{\alpha_L}}\right), \\
\label{equ:SFGLDE}
\end{equation}
where
$\alpha_{D}$ is the parameter of pure density evolution (hereafter ``PDE''; vertical shift of RLF), 
$\alpha_{L}$ is the parameter of pure luminosity evolution (hereafter ``PLE''; horizontal shift of RLF),  
and $\Phi_{0}^{\rm SFG}(L)$ is the local SFG RLF given by Eq. \ref{eqa:phi0}. 
If we assume that there is no evolution with redshift for $\Phi_\star$ and only $L_\star$ evolves with redshift ($\alpha_D = 0$),
we can define a SFG PLE model as
\begin{equation}
\Phi^{\rm SFG}_{\rm PLE}(L, z, \alpha_{L}) = \Phi_{0}^{\rm SFG}\left(\frac{L}{(1+z)^{\alpha_L}}\right).
\label{equ:SFGPLE}
\end{equation}
Similarly, SFG PDE model only allows $\Phi_\star$ to evolve with redshift ($\alpha_L = 0$),
which is defined as
\begin{equation}
\Phi^{\rm SFG}_{\rm PDE}(L, z, \alpha_{D}) = (1+z)^{\alpha_{D}}\times \Phi_{0}^{\rm SFG}(L).
\label{equ:SFGPDE}
\end{equation}

Then we used the Markov chain Monte Carlo (MCMC) algorithm in the Python package $\textsc{emcee}$ \citep{Foreman2013}
to fit the data (see Table \ref{tab:SFG}) with the above three models.
The averaged reduced $\chi^2$ over all the redshift bins of the best-fit LDE, PLE, and PDE models are
3.9, 4.8, and 526.4, respectively.
Firstly, we ignored the PDE model because it has the worst fit to the observational data.
The best-fit LDE and PLE models show significant differences at higher redshift ($z \gtrsim 1.5$)
where our data points only sample the bright end of RLF.
This will result in degeneracy in estimation of $\alpha_{\rm D}$ and $\alpha_{\rm L}$, 
preventing a precise calculation to the turnover position and turnover normalization.
Therefore, for simplicity, we only consider the PLE model (see Equation \ref{equ:SFGPLE}) here.

The best-fit $\alpha_L$ of the PLE model in each redshift bin is summarized in Table \ref{tab:RLFevolution}.
$\alpha_L$ shows an evolution with redshift,
which can be described by $\alpha_L = {\rm A}_L\times z+{\rm B}_L$.
We obtain the best-fit results with 
${\rm A}_L= -0.34\pm 0.01$ and ${\rm B}_L = 3.57\pm 0.01$ (see Panel A of Fig \ref{fig:LFevo}),
which are also summarized in Table \ref{tab:RLFevolution}.
The turnover position $L_{\star}$ is estimated by Equation \ref{equ:Lstar}, which is also listed in Table \ref{tab:RLFevolution}.
Our results are generally consistent with those in \cite{Novak2017} with the VLA-COSMOS 3 GHz project (see Fig. \ref{fig:LFtotal} and Fig. \ref{fig:LFevo}),
although our results show a slightly larger ${\rm B}_L$ (see Panel A of Fig. \ref{fig:LFevo}) 
and a slightly larger $L\star$ (see Panel B of Fig. \ref{fig:LFevo}), which may 
be due to the different selection criteria adopted for radio SFGs.
\cite{Novak2017} separated the entire radio sources sample into SFGs and radio-excess AGNs subsets,
while we firstly separate our entire radio sources sample into SFGs and QGs according to the UVJ selection criteria, and further distinguish the radio-excess AGNs from them.

\subsection{Fit the evolution of the AGN RLF}
\label{sec:RLFAGN}
The AGN RLF can be assumed as a double power-law shape \citep{Mauch2007}.
The local RLF of AGN \citep{Mauch2007} is defined as
\begin{equation}
\Phi_{0}^{\rm AGN} (L) = \frac{\Phi_{\star,0}}{(L_{\star,0}/L)^{\alpha}+(L_{\star,0}/L)^{\beta}},
\end{equation}
where $\Phi_{\star,0}=\frac{1}{0.5}10^{-5.5}\ \rm{Mpc}^{-3}\ \rm{dex}^{-1}$ is the turnover normalization of the local AGN RLF,
$L_{\star,0}=10^{24.59}\ \rm{W}\ \rm{Hz}^{-1}$ is the local turnover position,
$\alpha=-1.27$ and $\beta=-0.49$ are the indices at the bright and faint end, respectively.

Similar to the SFG RLF, we assume that AGN RLF has an unchanged shape at all cosmic times.
Thus, 
similar to Equations \ref{equ:SFGLDE} -- \ref{equ:SFGPDE},
LDE, PLE, and PDE models to describe AGN RLFs can be defined as
\begin{equation}
\Phi^{\rm AGN}_{\rm{LDE}}(L, z, \alpha_{L}, \alpha_{D}) = (1+z)^{\alpha_{D}}\times \Phi_{0}^{\rm AGN}\left(\frac{L}{(1+z)^{\alpha_L}}\right),
\label{equ:AGNLDE}
\end{equation}
\begin{equation}
\Phi^{\rm AGN}_{\rm{PLE}}(L, z, \alpha_{L}) = \Phi_{0}^{\rm AGN}\left(\frac{L}{(1+z)^{\alpha_L}}\right),
\label{equ:AGNPLE}
\end{equation}
\begin{equation}
\Phi^{\rm AGN}_{\rm{PDE}}(L, z, \alpha_{D}) = (1+z)^{\alpha_{D}}\times \Phi_{0}^{\rm AGN}(L),
\label{equ:AGNPDE}
\end{equation}
respectively. 
We also use the $\textsc{emcee}$ package to fit the observational data for radio-excess AGNs
(see Table \ref{tab:AGN}) with these three models.
The averaged reduced $\chi^2$ over all the redshift bins of the best-fit LDE, PDE, and PLE models are
2.3, 2.2, and 2.5, respectively.
Here we only consider the PDE model (see Equation \ref{equ:AGNPLE})
which has the best fit to the data. 

The best-fit $\alpha_{\rm D}$ of the PDE model in each redshift bin is
summarized in Table \ref{tab:RLFevolution},
which also shows a redshift evolution in a form of $\alpha_{\rm D}={\rm A}_D \times z + {\rm B}_D$.
The best-fit ${\rm A}_D$ and ${\rm B}_D$ are $-0.77\pm 0.06$ and $2.69\pm 0.09$, 
respectively (also see Table \ref{tab:RLFevolution}).
As we assumed an unchanged AGN RLF shape and a pure density evolution,
the evolution of $\Phi_{\star}$ (calculated by Equation \ref{equ:Phistar}) with redshift
can simply represent the number density distribution of AGN,
which peaked at $z \sim 1.5$ (see Panel D of Fig. \ref{fig:LFevo}).
Our result is generally consistent with that in \cite{Smolcic2017b} 
based on the VLA-COSMOS 3 GHz project (see Fig. \ref{fig:LFtotal} and Fig. \ref{fig:LFevo}),
and consistent with the AGN accretion rate density history obtained with X-ray surveys \citep[][and references therein]{Aird2010}.

We also study the RLF of the {\RAS} hosted by 
different galaxy populations (SFGs and QGs; see Fig. \ref{fig:RLFSFGQG}).
At $z < 1$, the radio-excess AGN population is mainly dominated by the sources hosted by QGs,
while radio-excess AGNs hosted by SFGs only have a comparable space densities at the faint end.
At $z > 1$, the space densities of radio-excess AGNs hosted by QGs decrease towards higher redshift, while
radio-excess AGNs hosted by SFGs show an opposite evolution trend.
In addition, radio-excess AGNs hosted by SFGs completely dominate 
those hosted by QGs at $z > 1.5$.
These trends are consistent with those for low-excitation radio galaxies (LERGs) in \cite{Kondapally2022}.
LERGs in \cite{Kondapally2022} are defined as objects showing powerful radio emission from AGN but not identified in other bands (e.g., IR and X-ray),
which represent 89\% of their entire radio-excess AGNs (11\% for high-excitation radio galaxies). For our radio-excess AGNs, nearly 80\% of objects do not have X-ray detections or have intrinsic 2--10 keV X-ray luminosity lower than $10^{42}\ \rm{erg}\ \rm{s}^{-1}$. Our radio-excess AGNs may be otherwise identified by IR or other methods, so LERG fraction in our {\RAS} is expected to be lower than 80\% while the fraction of high-excitation radio galaxies (HERGs) will be higher than 20\%.
In this work, our {\RAS} includes both LERGs and HERGs subsets, while \cite{Kondapally2022} mainly focused on LERGs.
Even so, both of their work and our work found that radio AGNs in SFGs and QGs have significantly different cosmic evolution.
Given that SFGs and QGs may have different fuelling mechanisms towards central SMBHs \citep[e.g.,][]{Kauffmann2009,Kondapally2022,Ni2023}, these results indicate that the radio activities of central engines
may depend on the fuelling mechanisms of their host galaxies.
In Section \ref{sec:Fagntotalsummary}, we will further investigate this topic through studying the probability of 
radio-excess AGNs hosted by different galaxy populations.

\begin{table*}[!t]
\centering
\setlength{\tabcolsep}{10pt}
\caption{Best-fit evolution parameters of radio Luminosity function for SFGs and radio-excess AGNs. \label{tab:RLFevolution}}
\begin{tabular}{cccccc}
\hline\hline\xrowht[()]{10pt}
 & \multicolumn{2}{c}{SFG (PLE)} & \multicolumn{2}{c}{Radio-excess AGN (PDE)} & \\
 \cmidrule(lr){2-3} \cmidrule(lr){4-5}
Redshift bin & $\alpha_{L}$ & $\log L_{\star}$ & $\alpha_{D}$ & $\log \phi_{\star}$ & $\log L_{\rm D}$ \\
 & & $[\rm{W}\ \rm{Hz}^{-1}]$ & & $[\rm{Mpc}^{-3}\ \rm{dex}^{-1}]$ & $[\rm{W}\ \rm{Hz}^{-1}]$ \\
\hline\xrowht[()]{5pt}
$0.1 < z \leq 0.5$  &  $3.37 \pm 0.21$  &  $21.65 \pm 0.02$  &  $1.88 \pm 0.52$  &  $-4.89 \pm 0.06$  &  $23.29 \pm 0.20$ \\
$0.5 < z \leq 1.0$  &  $3.48 \pm 0.06$  &  $22.11 \pm 0.01$  &  $2.36 \pm 0.15$  &  $-4.53 \pm 0.04$  &  $23.68 \pm 0.13$ \\
$1.0 < z \leq 1.5$  &  $3.08 \pm 0.04$  &  $22.35 \pm 0.02$  &  $1.34 \pm 0.11$  &  $-4.63 \pm 0.04$  &  $24.11 \pm 0.14$ \\
$1.5 < z \leq 2.0$  &  $2.92 \pm 0.04$  &  $22.55 \pm 0.02$  &  $1.30 \pm 0.11$  &  $-4.53 \pm 0.05$  &  $24.35 \pm 0.23$ \\
$2.0 < z \leq 3.0$  &  $2.70 \pm 0.02$  &  $22.74 \pm 0.01$  &  $0.83 \pm 0.09$  &  $-4.65 \pm 0.05$  &  $24.73 \pm 0.16$ \\
$3.0 < z \leq 4.0$  &  $2.42 \pm 0.03$  &  $22.85 \pm 0.02$  &  $0.58 \pm 0.15$  &  $-4.73 \pm 0.10$  &  $24.97 \pm 0.28$ \\
\hline\xrowht[()]{5pt}
Relation & \multicolumn{2}{c}{$L_{\star} = L_{\star,0}\times (1+z)^{\alpha_L}$} & \multicolumn{2}{c}{$\phi_{\star} = \phi_{\star,0}\times (1+z)^{\alpha_D}$} & $L_{\rm D} = L_{\rm D,0}\times(1+z)^{\alpha_{\rm LD}}$ \\
 & \multicolumn{2}{c}{$\alpha_L = {\rm A}_L\times z + {\rm B}_L$} & \multicolumn{2}{c}{$\alpha_D = {\rm A}_D\times z + {\rm B}_D$} & $\alpha_{\rm LD}={{\rm C}\times z + {\rm D}}$ \\
 \hline\xrowht[()]{5pt}
Best-fit parameters & \multicolumn{2}{c}{${\rm A}_L = -0.34\pm 0.01$} & \multicolumn{2}{c}{${\rm A}_D = -0.77\pm 0.06$} & ${\rm C} = 0.06\pm 0.01$ \\
 & \multicolumn{2}{c}{${\rm B}_L = 3.57\pm 0.01$} & \multicolumn{2}{c}{${\rm B}_D = 2.69\pm 0.09$} & ${\rm D} = 3.05\pm 0.02$ \\
\hline
\end{tabular}
\tablefoot{
For SFGs, $L_{\star,0} = 1.85\times10^{21}\ \rm{W}\ \rm{Hz}^{-1}$, 
represents the turnover position of the local SFG RLF \citep{Novak2017}.
For radio-excess AGNs, $\Phi_{\star,0}=\frac{1}{0.5}10^{-5.5}\ \rm{Mpc}^{-3}\ \rm{dex}^{-1}$,
represents the turnover normalization of the local AGN RLF \citep{Mauch2007}.
The local turnover luminosity $L_{\rm D,0} = 10^{22.9}\ {\rm W}\ {\rm Hz}^{-1}$, is obtained by the cross point between
the local SFG RLF \citep{Novak2017} and the local AGN RLF \citep{Mauch2007}.
}
\end{table*}

\subsection{Crossover luminosity between SFG RLF and AGN RLF}
\label{sec:turnoverL}
The comparison between the SFG RLF and AGN RLF shows that 
AGNs will dominate the radio populations beyond a certain luminosity
(hereafter turnover luminosity $L_{\rm D}$; see the magenta dash-dotted lines in Fig. \ref{fig:LFtotal}).
In the local universe, this crossover luminosity is around $10^{23}\ \rm{W}\ \rm{Hz}^{-1}$,
which is usually used to select RL AGNs \citep[e.g.,][]{Best2005,Kukreti2023}.
Towards higher redshift, given that both SFG RLFs and AGN RLFs show evolutions with redshift,
this crossover luminosity will not be expected to be constant at all cosmic time.
Our results indicate that the redshift evolution of this crossover luminosity can be described by
\begin{equation}
L_{\rm D} = L_{\rm D,0} \times (1+z)^{\alpha_{LD}},
\label{equ:LD}
\end{equation}
where $L_{\rm D,0}$ is the crossover luminosity in the local universe,
$\alpha_{LD}$ is the evolution index of the crossover luminosity.
Here, $L_{\rm D,0} = 10^{22.9}\ {\rm W}\ {\rm Hz}^{-1}$, is obtained by the cross point between
the local SFG RLF \citep{Novak2017} and local AGN RLF \citep{Mauch2007}.
Further, $\alpha_{LD}$ also shows a weak evolution with redshift:
\begin{equation}
\alpha_{LD} = (0.06\pm 0.01)\times z + (3.05\pm 0.02).
\label{equ:aLD}
\end{equation}
These results provide us another way to 
select powerful radio AGNs at different redshift through solely radio survey. 
It means that if IR data are not available, a radio source at redshift $z$ can be selected as a radio AGN when
its 1.4 GHz radio luminosity is larger than 
$L_{\rm D} = 10^{22.9} \times (1+z)^{0.06\times z + 3.05}\ {\rm W}\ {\rm Hz}^{-1}$.
This crossover luminosity describes a luminosity threshold where radio-AGNs begin to dominate
entire radio population. This result is also consistent with those from previous works \citep[e.g.,][]{McAlpine2013},
and also consistent with the radio luminosity threshold proposed for radio-AGN selection in the past \citep[e.g.,][]{Magliocchetti2017}.

\section{The probability of hosting a radio-excess AGN in SFGs and QGs}
\label{sec:Fagntotalsummary}

In this section, we aim to study the probability of different galaxy populations hosting a radio-excess AGN.
Here we use the combined \texttt{All Galaxies Sample} (see Section \ref{sec:data} and Fig. \ref{fig:flowchart}) 
and combined \texttt{Radio-excess AGN Sample} (see Section \ref{sec:REAGNpaper} and Fig. \ref{fig:flowchart})
in the GOODS-N and COSMOS/UltraVISTA fields.
The \texttt{All Galaxies Sample} and the host galaxies of the \texttt{Radio-excess AGN Sample}
had been divided into SFGs and QGs according to the UVJ selection method (see Section \ref{sec:UVJ}).
In Sections \ref{sec:Fagnobsdata} and \ref{sec:chi2}, 
we use the observational data to calculate the probability of 
a SFG or a QG with the stellar mass $M_{\star}$ and at redshift $z$ hosting a
radio-excess AGN with the 1.4 GHz luminosity $L_{\rm R}$ in each redshift bin.
In Section \ref{sec:pzevl}, we use the method of \cite{Aird2012} to quantitatively
calculate the probability of a SFG or a QG hosting a radio-excess AGN as
a function of stellar mass, radio luminosity, and redshift.
This method is based on the maximum-likelihood approach,
which does not require data binning for stellar mass, luminosity, and redshift \citep{Aird2012}. 
Here we do not consider the star-formation contamination for the AGN radio luminosity,
while for most of our {\RAS}, contribution from star-formation to the radio luminosity have been verified to be not significant compared to that from AGN.

\subsection{Calculate the probability for observational data}
\label{sec:Fagnobsdata}
In order to conveniently compare the data and the best-fit model,
we divided our {\TGS} and {\RAS} into 6 redshift bins 
(same as the redshift bins in Section \ref{sec:REAGNpaper}).
Within each redshift bin, we subdivided our {\TGS} and 
{\RAS} into different stellar mass bins.
The conditional probability density function $p(L_{\rm R}\ |\ M_{\star},\ z)$ 
describes the probability of a galaxy with stellar mass $M_{\star}$ and at redshift $z$ hosting a
radio-excess AGN with 1.4 GHz luminosity $L_{\rm R}$.
Following \cite{Aird2012}, here $p(L_{\rm R}\ |\ M_{\star},\ z)$ is defined as the probability density
per logarithmic luminosity interval (units are dex$^{-1}$).
Thus, $p(L_{\rm R}\ |\ M_{\star},\ z)$
can be converted to AGN fraction ($f_{\rm AGN}$) according to
\begin{equation}
f_{\rm AGN}(M_{\star},\ z) = \int_{{\log L_{\rm lim}}}^{\infty}p(L_{\rm R}\ |\ M_{\star},\ z)\ d \log L_{\rm R}.
\end{equation}
Here $f_{\rm AGN}$ is defined as the fraction of galaxies with stellar mass $M_{\star}$
and at redshift $z$ that host a radio-excess AGN,
$L_{\rm lim}$ is the lower limit of the radio luminosity.
For the combined observational data from the GOODS-N and COSMOS/UltraVISTA fields, in each redshift bin,
$p(L_{\rm R}\ |\ M_{\star},\ z)$ in the $m$th stellar mass bin and the $n$th luminosity bin is defined by
\begin{equation}
p(L_{n}\ |\ M_{m},\ z) = \frac{N_{{\rm AGN},mn}}{N_{{\rm gal},m}\Delta \log L_{n}} = \frac{f_{\rm AGN}(M_{\star},\ z)}{\Delta \log L_{n}},
\label{eq:Fagndata}
\end{equation}
where $N_{{\rm AGN},mn}$ is the number of the {\RAS} 
in the $m$th stellar mass bin and the $n$th luminosity bin,
$N_{{\rm gal},m}$ is the number of the {\TGS} in the $m$th stellar mass bin,
and $\Delta \log L_{n}$ is the width of the $n$th luminosity bin.
It is worth stating that when calculating the probability of SFGs (or QGs, or all galaxies) 
hosting a radio-excess AGN, here $N_{{\rm AGN},mn}$ refers to all 
the radio-excess AGNs hosted by SFGs (or QGs, or all galaxies)
and
$N_{\rm gal}$ refers to all the SFGs (or QGs, or all galaxies) in the {\TGS}.
The estimates of $p(L_{\rm R}\ |\ M_{\star},\ z)$ for SFGs in 6 redshift bins
are shown as colored symbols in Fig. \ref{fig:FagnSFGdatawithmass} and Fig. \ref{fig:FagnSFGdatawithLR}.
The radio-excess AGNs in both QGs and all galaxies exhibit similar trends to SFGs,
so we do not show their details here.
The probability of a galaxy hosting a radio-excess AGN with a given $L_{\rm R}$
increases with $M_\star$ at all redshift bins (see Fig. \ref{fig:FagnSFGdatawithmass}), 
which is consistent with the previous works \citep[e.g.,][]{Best2005,Sabater2019}.
The probability of a galaxy with a given $M_\star$ hosting a radio-excess AGN decreases
with $L_{\rm R}$, which is also consistent with previous works \citep[e.g.,][]{Best2005}.
These trends of radio-excess AGNs in this work are consistent with those of X-ray AGNs \citep[e.g.,][]{Haggard2010,Aird2012,Wang2017}.

\subsection{Simple $\chi^2$ fits in each redshift bin}
\label{sec:chi2}
To quantitatively study the trends of $p(L_{\rm R}\ |\ M_{\star},\ z)$,
we apply $\chi^2$ fits to the results calculated by Equation \ref{eq:Fagndata} 
(see the data points in Fig. \ref{fig:FagnSFGdatawithmass} and Fig. \ref{fig:FagnSFGdatawithLR}).
Next we take the analysis for SFGs as an example to show the detailed analysis procedures.
At each fixed $L_{\rm R}$,
we assume a simple power-law relation for $p(L_{\rm R}\ |\ M_{\star},\ z)$ as
a function of $M_{\star}$,
\begin{equation}
\log[p(L_{\rm R}\ |\ M_{\star},\ z)] = a + b \log\left[\frac{M_{\star}}{10^{11}\ M_{\odot}}\right],
\label{eq:FagnLR}
\end{equation}
where $a$ and $b$ are the intercept and slope of the relation, respectively.
The best-fit values for $a$ and $b$ in each $L_{\rm R}$ bin and 
each redshift bin are presented in the right column of Fig. \ref{fig:FagnSFGdatawithmass}.
The averaged best-fit reduced-$\chi^2$ over all the redshift bins and all the $L_{\rm R}$ bins is 1.22.
In all redshift bins, the intercept $a$ decreases with larger $L_{\rm R}$, 
while the slope $b$ does not show significant changes.
Similarly, for each fixed $M_\star$, $p(L_{\rm R}\ |\ M_{\star},\ z)$ as a function of $L_{\rm R}$ can be defined as
\begin{equation}
\log[p(L_{\rm R}\ |\ M_{\star},\ z)] = c + d \log\left[\frac{L_{\rm R}}{10^{23}\ \rm{W}\ \rm{Hz}^{-1}}\right],
\label{eq:Fagnmass}
\end{equation}
where $c$ and $d$ are the intercept and slope of the relation, respectively.
The best-fit parameters in each $M_\star$ bin and each redshift bin are 
shown in the right column of Fig. \ref{fig:FagnSFGdatawithLR}.
The averaged best-fit reduced-$\chi^2$ over all the redshift bins and all the $M_\star$ bins is 1.76.
In all redshift bins, the intercept $c$ decreases with higher $M_\star$, while the slope $d$ nearly keeps constant.
These results indicate that
the slope $b$ (or the slope $d$) 
of $p(L_{\rm R}\ |\ M_{\star},\ z)$ as a function of 
$M_\star$ (or $L_{\rm R}$) is independent of $L_{\rm R}$ (or $M_\star$).
QGs and all galaxies show similar trends but different best-fit parameters comparing to SFGs.
Given that we mainly focus on the trends in this section, for brevity,
we do not discuss their results in details here.
The main differences among different galaxy populations are 
discussed in Section \ref{sec:FagnSFGQG} and shown in Fig. \ref{fig:Fagnevo}.

\subsection{Redshift evolution}
\label{sec:pzevl}
The above analysis starts by dividing sources into different redshift bins, 
then binning the sources according to their $M_{\star}$ and $L_{\rm R}$ within each redshift bin,
and finally fitting the binned data.
Therefore, the best-fit result may be significantly affected by the bin size of redshift, $M_{\star}$, and $L_{\rm R}$,
and the information of each individual galaxy cannot be fully utilized due to the binning.
Therefore, next we utilize the maximum likelihood
fitting approach in \cite{Aird2012} to measure the dependence of
$p(L_{\rm R}\ |\ M_{\star},\ z)$ on both $M_{\star}$ and $L_{\rm R}$.
At first, in Section \ref{sec:maximum}, we introduce the fitting approach applied to our samples.
Then we test this approach in each redshift bin in Section \ref{sec:Fagnmaxresults}.
Finally, in Section \ref{sec:maximumz}, we apply this approach incorporating the redshift evolution in order to
remove the effects brought by the redshift binning.

\subsubsection{Maximum-likelihood fitting}
\label{sec:maximum}
According to the results in Section \ref{sec:chi2},
the slope $b$ (or $d$) in Equation \ref{eq:FagnLR} (or \ref{eq:Fagnmass})
of $p(L_{\rm R}\ |\ M_{\star},\ z)$ as a function of 
$M_{\star}$ (or $L_{\rm R}$) is independent of $L_{\rm R}$ (or $M_{\star}$).
It means that $p(L_{\rm R}\ |\ M_{\star},\ z)$ at a fixed redshift can be expressed
as a separable function of $M_{\star}$ and $L_{\rm R}$ in the form of
\begin{equation}
\label{eq:pLMnoz}
p(L_{\rm R}\ |\ M_{\star},\ z)\ d\log L_{\rm R} = K\left(\frac{M_{\star}}{M_0}\right)^{\gamma_M}\left(\frac{L_{\rm R}}{L_0}\right)^{\gamma_L} d \log L_{\rm R},\\
\end{equation}
where the scaling factor $M_0$ is set to $10^{11}\ M_\odot$, the scaling factor $L_0$ is set to $10^{23}\ {\rm W}\ {\rm Hz}^{-1}$,
$K$ is the normalization, $\gamma_M$ and $\gamma_L$ are indices.
The best-fit parameters are found through maximizing the
log-likelihood function,
\begin{equation}
\label{eq:likelihoodF}
\ln \mathcal{L} = - \mathcal{N} + \sum_{k = 1}^{N_i^{\rm AGN}} \ln p_k,
\end{equation}
where $N_i^{\rm AGN}$ is the number of radio-excess AGNs in 
the GOODS-N and COSMOS/UltraVISTA fields in the $i$th redshift bin,\
$p_k$ is the probability of a galaxy with stellar mass $M_{k}$ hosting
a radio-excess AGN with 1.4 GHz luminosity $L_{k}$:
\begin{equation}
\label{eq:pk}
p_k = p(L_k\ |\ M_k,\ z_k) = K\left(\frac{M_k}{M_0}\right)^{\gamma_M}\left(\frac{L_k}{L_0}\right)^{\gamma_L}.
\end{equation}
The expected number of radio-excess AGNs, $\mathcal{N}$, is defined as
\begin{equation}
\label{eq:expectN}
\begin{aligned}
\mathcal{N} &= \sum_{j = 1}^{N_i^{\rm gal,gn}} \int_{\log L_{{\rm lim,gn},j}}^{\infty} p(L_{\rm R}\ |\ M_j,\ z_j)\ d \log L_{\rm R} \\
&+ \sum_{j = 1}^{N_i^{\rm gal,cs}} \int_{\log L_{{\rm lim,cs},j}}^{\infty} p(L_{\rm R}\ |\ M_j,\ z_j)\ d \log L_{\rm R} \\
&= \sum_{j = 1}^{N_i^{\rm gal,gn}} K \left(\frac{M_j}{M_0}\right)^{\gamma_M} \int_{\log L_{{\rm lim,gn},j}}^{\infty} \left(\frac{L_{\rm R}}{L_0}\right)^{\gamma_L}\ d\log L_{\rm R}, \\
&+ \sum_{j = 1}^{N_i^{\rm gal,cs}} K \left(\frac{M_j}{M_0}\right)^{\gamma_M} \int_{\log L_{{\rm lim,cs},j}}^{\infty} \left(\frac{L_{\rm R}}{L_0}\right)^{\gamma_L}\ d\log L_{\rm R},
\end{aligned}
\end{equation}
where $N_i^{\rm gal,gn}$ and $N_i^{\rm gal,cs}$ are the numbers of galaxies 
in the $i$th redshift bin for the GOODS-N and COSMOS/UltraVISTA fields, respectively.
$L_{\rm lim}$ of the GOODS-N field ($L_{{\rm lim,gn},j}$) corresponds to the $5\sigma$ 
detection limit of the radio survey at redshift $z_j$ (see Fig. \ref{fig:radiolimit}),
while for the COSMOS ($L_{{\rm lim,cs},j}$) field,
it corresponds to the $L_{\rm 1.4 GHz}$ threshold for correcting the AGN 
classification purity (see details
in Appendix \ref{sec:REAGNpurity}).
The $1\sigma$ error of each model parameter is estimated by the maximum projection 
of $\Delta \mathcal{S} = 1.0$ where $\mathcal{S} = -2\ln \mathcal{L}$ according to \cite{Aird2012}.

\subsubsection{Results in each redshift bin}
\label{sec:Fagnmaxresults}
In each redshift bin,
we use the above maximum-likelihood fitting approach to constrain $p(L_{\rm R}\ |\ M_{\star},\ z)$
for SFGs, QGs, and all galaxies.
The best-fit values for $\gamma_M$, $\gamma_L$, and $K$ in each redshift bin
are shown in the form of colored symbols in Fig. \ref{fig:Fagnevo}.
For both SFGs and QGs, we find that $\gamma_M$ and $\gamma_L$ 
almost keep constant at all redshift bins 
except at $z \lesssim 0.5$ (see detailed discussions in Section \ref{sec:maximumz}).
For all galaxies, $\gamma_M$ and $\gamma_L$ do not show any significant changes with redshift ($0.1 < z \leq 4.0$),
which is consistent with the trends of X-ray AGNs in \cite{Aird2012} at $0.2 < z \leq 1.0$.
For SFGs, QGs and all galaxies, the normalization $K$ significantly increases with redshift.
In addition, SFGs and QGs show significantly different $\gamma_M$, $\gamma_L$, and $K$,
which will be discussed in details in Section \ref{sec:FagnSFGQG}.
Overall, the probability of a galaxy with a given stellar mass hosting
a radio-excess AGN with a given radio luminosity increases towards higher redshift.
This indicates higher AGN activities and potentially more significant AGN feedbacks towards higher redshift.

\subsubsection{Incorporating redshift evolution into the maximum-likelihood fitting}
\label{sec:maximumz}
\begin{figure}[t!]
\centering
\includegraphics[width=\linewidth, clip]{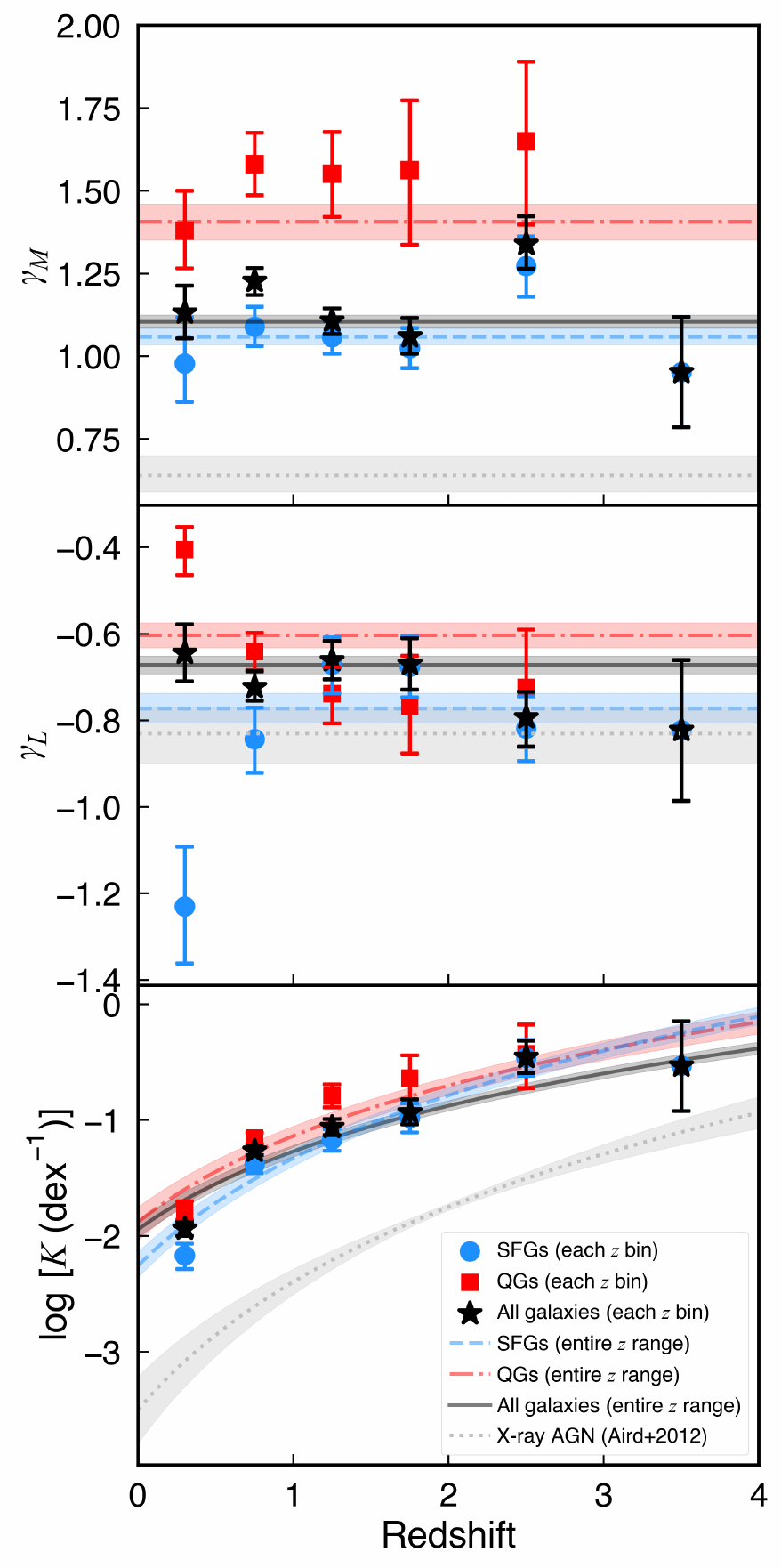}
\caption{Best-fit parameters of the maximum-likelihood fitting for 
$p(L_{\rm R}\ |\ M_{\star},\ z)$ of SFGs 
(blue circles and blue dashed lines), QGs (red squares and red dash-dotted lines), 
and all galaxies (black stars and black solid lines).
Colored symbols represent the best-fit parameters obtained by 
Equations \ref{eq:pLMnoz}--\ref{eq:expectN} in each redshift bin
(see details in Sections \ref{sec:maximum} and \ref{sec:Fagnmaxresults}).
The colored lines show the best-fit results over the entire redshift range
obtained by Equations \ref{eq:pLMz}--\ref{eq:Knormz} and \ref{eq:likelihoodF}--\ref{eq:expectN}
(see details in Section \ref{sec:maximumz}).
The dotted gray lines represent the results for the X-ray AGNs in \cite{Aird2012}
\label{fig:Fagnevo}}
\end{figure}

According to the results of Section \ref{sec:Fagnmaxresults},
both $\gamma_M$ and $\gamma_L$ are nearly constant over the entire redshift range,
which indicates that the redshift evolution of $p(L_{\rm R}\ |\ M_{\star},\ z)$ 
can be assumed to be independent of $L_{\rm R}$ and $M_{\star}$.
Only the normalization $K$ shows an evident evolution with redshift, 
so $p(L_{\rm R}\ |\ M_{\star},\ z)$ incorporating the redshift evolution for $K$ with a simple power-law can be rewritten as 
\begin{equation}
\label{eq:pLMz}
p(L_{\rm R}\ |\ M_{\star},\ z)\ d\log L_{\rm R} 
= A\left(\frac{1+z}{1+z_0}\right)^{\gamma_z} \left(\frac{M_{\star}}{M_0}\right)^{\gamma_M}\left(\frac{L_{\rm R}}{L_0}\right)^{\gamma_L} \ d \log L_{\rm R},
\end{equation}
where the scaling factor $z_0$ is set to be 2.0 (median of the redshift range for our sample),
$A$ is the overall normalization.
Thus,
\begin{equation}
\label{eq:Knormz}
K = A\left(\frac{1+z}{1+z_0}\right)^{\gamma_z}.
\end{equation}
Then we modify $p(L_{\rm R}\ |\ M_{\star},\ z)$ in Equations \ref{eq:likelihoodF}--\ref{eq:expectN}
with Equation \ref{eq:pLMz} to perform maximum-likelihood fitting over the entire redshift range ($0.1 < z \leq 4.0$).
The best-fit parameters are listed in Table \ref{tab:Fagnz} and plotted in Fig. \ref{fig:Fagnevo} as the colored lines.
The redshift evolution model over the entire redshift range
is almost consistent with the result obtained in each redshift bin (colored symbols in Fig. \ref{fig:Fagnevo}).

\begin{figure*}[t!]
\centering
\includegraphics[width=\linewidth, clip]{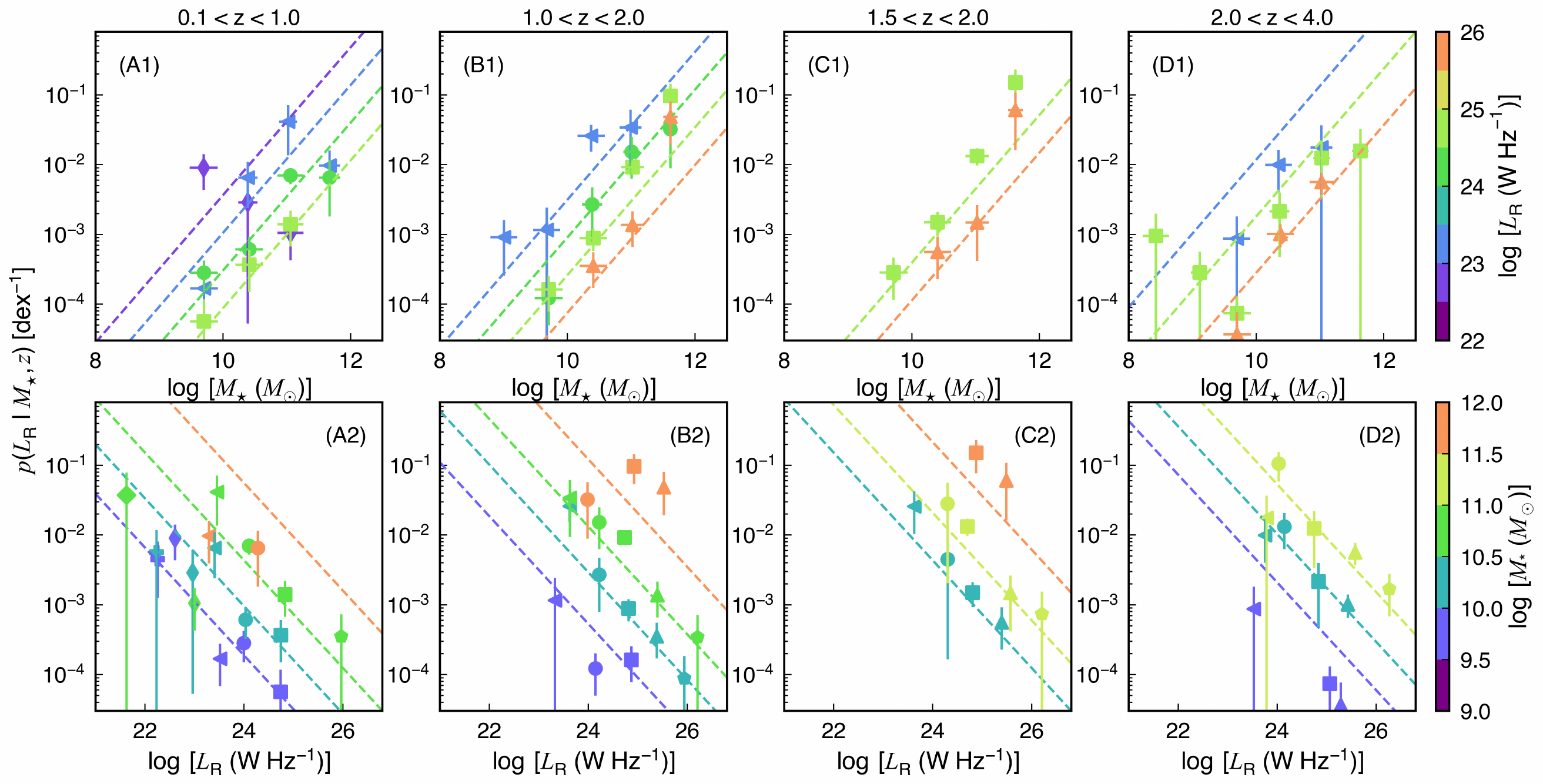}
\caption{Estimates of $p(L_{\rm R}\ |\ M_{\star},\ z)$
as a function of $M_{\star}$ and $L_{\rm R}$
based on our maximum-likelihood fitting results for SFGs.
Colored dashed lines in both top and bottom rows
represent our best-fit model from the unbinned maximum-likelihood fitting (see Equation \ref{eq:pLMz} and Table \ref{tab:Fagnz})
through combining data in the GOODS-N and COSMOS/UltraVISTA fields
over the entire redshift range ($0.1 < z < 4$) evaluated at the center of each redshift bin.
Binned data points are scaled with the probability estimated by the model (see Equation \ref{eq:pLMz})
using the $N_{\rm obs}/N_{\rm mdl}$ method of \cite{Aird2012} (see details in Section \ref{sec:maximumz}).
In the top panel, different colors represent different $M_{\star}$ bins.
In the bottom panel, different colors show different $L_{\rm R}$ bins.
\label{fig:FagnSFGmo}}
\end{figure*}

To compare the model with the observational data, 
we use the $N_{\rm obs}/N_{\rm mdl}$ method of \cite{Aird2012} (see Equation 14 in their paper)
to scale the observed radio-excess AGN number ($N_{\rm obs}$) 
by the expected number estimated by the model ($N_{\rm mdl}$) in each redshift bin (see Equation \ref{eq:pLMz}).
Overall, the binned estimates for the observational data in each redshift bin are well described 
by the best-fit model obtained with maximum-likelihood fitting over the entire redshift range 
(see results for SFGs in Fig. \ref{fig:FagnSFGmo}, for QGs in Fig. \ref{fig:FagnQGmo},
for all galaxies in Fig. \ref{fig:FagnTotmo}).
However, the best-fit model cannot fully explain the data in the following cases:
low-luminosity radio-excess AGN in SFGs at $0.1 < z \leq 1.0$ 
($L_{\rm 1.4GHz} < 10^{23}\ \rm{W}\ \rm{Hz}^{-1}$; see Panel A1 in Fig. \ref{fig:FagnSFGmo})
and at $2 < z \leq 4$
($L_{\rm 1.4GHz} < 10^{23.5}\ \rm{W}\ \rm{Hz}^{-1}$; see Panel D1 in Fig. \ref{fig:FagnSFGmo}), 
radio-excess AGN in massive SFGs ($M_{\star} > 10^{11.5}\ M_\odot$) 
at $0.1 < z \leq 1.0$ (see Panels A2 in Fig. \ref{fig:FagnSFGmo}), 
radio-excess AGN in low-mass QGs ($M_{\star} < 10^{10}\ M_\odot$) 
at $0.1 < z \leq 1.0$ (see Panels A2 in Fig. \ref{fig:FagnQGmo}, repectively).
The following reasons might explain the slight differences between observational data and model predictions.
In the analysis, we assume that $\gamma_M$ (or $\gamma_L$) is 
independent of $L_{\rm R}$ (or $M_\star$) and redshift 
(see Sections \ref{sec:chi2} and \ref{sec:Fagnmaxresults}).
On the one hand, a mild dependence of $\gamma_M$ (or $\gamma_L$) 
on $L_{\rm R}$ (or $M_\star$) cannot be fully ruled out. 
In addition, $\gamma_M$ may be different between low-mass and high-mass galaxies \citep{Williams2015,Zhu2023},
while $\gamma_L$ may also change from low-luminosity to high-luminosity populations \citep{Best2005}.
On the other hand, a weak evolution of both $\gamma_M$ and $\gamma_L$ 
with redshift might exist (see colored symbols in Fig. \ref{fig:Fagnevo} and discussion in Section \ref{sec:Mqir}).
Some works had found that $\gamma_{M}$ decreases 
from the local universe to $z \sim 2$ \citep{Williams2015,Zhu2023},
while \cite{Kondapally2022} found a weak positive evolution with redshift (for SFGs only).
Even so, on the whole our model can still well describe the 
redshift evolution of radio-excess AGN fraction across the entire redshift range.

\begin{table}[t]
\centering
\setlength{\tabcolsep}{3pt}
\caption{Best-fit parameters from Maximum-likelihood Fitting with redshift evolution. \label{tab:Fagnz}}
\begin{tabular}{cccccl}
\hline\hline\xrowht[()]{10pt}
Type & $\log A$ & $\gamma_M$ & $\gamma_L$ & $\gamma_z$  \\
\hline\xrowht[()]{5pt}
SFG & $-0.79\pm 0.07$ & $1.06\pm 0.03$ & $-0.77\pm 0.03$ & $3.08\pm 0.20$  \\
QG & $-0.70\pm 0.08$ & $1.41\pm 0.05$ & $-0.60\pm 0.03$ & $2.47\pm 0.28$  \\
All & $-0.88\pm 0.05$ & $1.10\pm 0.02$ & $-0.67\pm 0.02$ & $2.24\pm 0.14$  \\
\hline
\end{tabular}
\end{table}

\subsubsection{Quantitative relation of radio-excess AGN fraction in SFGs and QGs}
\label{sec:FagnSFGQG}
To sum up, we obtain quantitative relations for the probability of SFGs, QGs, or all galaxies hosting a radio-excess AGN 
as a function of $M_\star$, $L_{\rm R}$, and redshift, respectively:
\begin{equation}
\begin{aligned}
p(L_{\rm R}\ |\ M_{\star},\ z)_{\rm SFG} 
&= 10^{-0.79} \left(\frac{M_{\star}}{M_0}\right)^{1.06} \left(\frac{L_{\rm R}}{L_0}\right)^{-0.77} \left(\frac{1+z}{1+z_0}\right)^{3.08}, \\
p(L_{\rm R}\ |\ M_{\star},\ z)_{\rm QG} 
&= 10^{-0.70} \left(\frac{M_{\star}}{M_0}\right)^{1.41} \left(\frac{L_{\rm R}}{L_0}\right)^{-0.60} \left(\frac{1+z}{1+z_0}\right)^{2.47}, \\
p(L_{\rm R}\ |\ M_{\star},\ z)_{\rm All} 
&= 10^{-0.88} \left(\frac{M_{\star}}{M_0}\right)^{1.10} \left(\frac{L_{\rm R}}{L_0}\right)^{-0.67}\ \left(\frac{1+z}{1+z_0}\right)^{2.24},
\label{eq:Fagnreal}
\end{aligned}
\end{equation}
where $M_0$ is set to $10^{11}\ M_\odot$, $L_0$ is set to $10^{23}\ {\rm W}\ {\rm Hz}^{-1}$,
and $z_0$ is set to 2.0. These best-fit parameters and their uncertainties are summarized in Table \ref{tab:Fagnz}.
SFGs and QGs have 
significantly different evolution trends for radio-excess AGN fraction (also see Fig. \ref{fig:Fagnevo}).
QGs have larger $\gamma_M$ and $\gamma_L$ than SFGs, which indicates
that radio-excess AGNs in QGs prefer to reside in more massive galaxies 
with higher radio luminosity compared to those in SFGs.
Both SFGs and QGs have an increasing radio-excess AGN fraction with redshift (see Fig. \ref{fig:Fagnevo}).
SFGs show a larger $\gamma_z$ than QGs (see Table \ref{tab:Fagnz}), which implies that
radio-excess AGN fractions in SFGs increase more rapidly towards higher redshift 
than those in QGs.

\section{Discussion: comparison with previous works}
\label{sec:Totdis}

\label{sec:Mqir}
As Equation \ref{eq:Fagnreal} shows, we obtain a smaller $\gamma_{M}$ ($\sim 1$)
than that in \cite{Best2005} ($\gamma_{M}\sim 2.5$; $0.03 < z < 0.3$).
Our result is consistent with that in \cite{Williams2015} at $1.5 < z < 2$ ($\gamma_{M}\sim 1.0$), while their results
showed a decreasing $\gamma_{M}$ trend from the local universe ($\gamma_{M}\sim 2.7$) to $z \sim 2$.
\cite{Zhu2023} also found a decreasing stellar-mass dependence from $z \sim 0.3$ to  $z \sim 2.3$.
In addition, the dependence of radio-AGN fraction on the stellar mass may change when $M_* > 10^{11} M_\odot$
\citep{Williams2015,Zhu2023}.
The differences between our and their results may be due to the 
different radio-AGN selections and different calculation methods.
Previous works usually focused on powerful radio-AGNs ($L_{\rm 1.4 GHz} > 10^{23}\ \rm{or}\ 10^{24}\ \rm{W}\ \rm{Hz}^{-1}$),
while our sample additionally includes some faint radio-AGNs with $L_{\rm 1.4 GHz} < 10^{23}\ \rm{W}\ \rm{Hz}^{-1}$.
Moreover, the different evolution trends of radio-AGN in SFGs and QGs indicate that 
disregarding the galaxy types may have a non-negligible effect on statistical results.
\cite{Kondapally2022} proposed for the first time a radio-AGNs study for QGs and SFGs separately.
They found that low-excitation radio galaxies in QGs have a larger 
$\gamma_{M}$ ($\sim 2.5$) than those in SFGs ($\gamma_{M}\sim 1.37$).
We found a similar trend but lower $\gamma_{M}$ values than their results 
($\gamma_{M}\sim 1.4$ for QGs and $\gamma_{M}\sim 1.0$ for SFGs in our work).
One possible reason for this difference may be that we focus on 
different radio-AGNs samples comparing to \cite{Kondapally2022} (see details in Section \ref{sec:RLFAGN}).
Our results show that $\gamma_{M}$ has no significant evolution with redshift at $0.1 < z \leq 4.0$,
while for LERGs in \cite{Kondapally2022} at $0.3 < z \leq 1.5$, $\gamma_{M}$ shows no significant evolution for QGs and a positive evolution for SFGs.

Given that we use the same calculation method as \cite{Aird2012} (see details in Section \ref{sec:pzevl}), 
we also compare the results for radio-excess AGNs with those for X-ray AGNs from \cite{Aird2012}
which are plotted as gray dotted lines in Fig. \ref{fig:Fagnevo}.
Comparing to X-ray AGNs \citep{Aird2012}, evolutions of radio-excess AGN fraction 
(see black solid lines in Fig. \ref{fig:Fagnevo}) have a 
larger $\gamma_M$, which indicates that radio-excess AGNs 
tend to reside in more massive galaxies (see Fig. \ref{fig:Fagnevo}).
X-ray AGN fractions in red sequence galaxies exhibit a more 
rapid redshift evolution than those in blue cloud galaxies \citep{Aird2012,Wang2017},
while radio-excess AGNs fraction in SFGs
increases more rapidly towards higher redshift than those in QGs.
These different evolution trends between radio-excess AGNs and X-ray AGNs suggest
that black holes with different accretion states may influence their host galaxies in different modes.

In addition, we also examine whether $M_\star$-dependent IRRC \citep{Delvecchio2021}
affects the results about the probability of hosting a radio-excess AGN.
Following \cite{vanderVlugt2022}, we select radio-excess AGNs
with $q_{\rm TIR}$ deviating more than $3\sigma$
from the $M_\star$-dependent IRRC of \cite{Delvecchio2021}. 
Then we recalculate the probability of a radio-excess AGN hosted by SFGs
using Equation \ref{eq:pLMz}
(see the dark blue region in Fig. \ref{fig:Fagnevocompare})
which is consistent with the results based on the radio-excess-AGN selection method in Section \ref{sec:REAGNpaper} 
(see the green region in Fig. \ref{fig:Fagnevocompare}).
This result indicates that using the $M_*$-dependent IRRC \citep{Delvecchio2021}
does not alter the results in this work.

\section{Summary}
\label{sec:summary}
Firstly we use the optical to MIR surveys to select a {\TGS} 
with $\sim$ 400,000 sources at $0.1 < z < 4$ in 
GOODS-N and COSMOS/UltraVISTA fields (totaling 1.6 degree$^2$; Section \ref{sec:data}).
After cross-matching with the deep/large radio surveys (1.4 GHz or 3 GHz) in these fields,
we select 7494 radio sources with S/N of radio flux $\geq 5$ at $0.1 < z < 4$ as our {\RSS} (Section \ref{sec:data}).
Combining with the de-blended IR photometry in these fields,
we calculate the infrared-radio ratio ($q_{\rm TIR}$) distributions of these radio sources
to select a {\RAS} with 983 sources (Section \ref{sec:qTIRagn}).
Next we subdivide all our samples into SFGs and QGs using the UVJ method.
Based on the {\RSS}, we construct 1.4 GHz radio luminosity functions (RLFs) for SFGs
and radio-excess AGNs at $0.1 < z < 4$ (Section \ref{sec:RLFobs}), respectively,
and study their evolutions with redshift.
Based on the {\TGS} and {\RAS}, we further investigate the probability of different galaxies populations (SFGs and QGs)
hosting a radio-excess AGN as a function
of stellar mass, radio luminosity, and redshift.
The main conclusions we have obtained are shown as follows.

\begin{enumerate}
  \item The $q_{\rm TIR}$ value of SFGs shows a weak evolution with redshift: 
$\bar{q}_{\rm{TIR}} = 2.62\times (1+z)^{-0.08}$ (Section \ref{sec:qIRevolution}),
which is generally consistent with previous works.
  \item The evolution of RLFs with redshift for SFGs
can be well described by the pure luminosity evolution model of 
$L_{\star}\propto (1+z)^{-0.34\times z +3.57}$ (Section \ref{sec:RLFSFG}).
The evolution of AGN RLFs follows the pure density evolution model of 
$\Phi_{\star} \propto (1+z)^{-0.77\times z +2.69}$ (Section \ref{sec:RLFAGN}).
  \item The evolution of crossover luminosity between SFG RLFs and AGN RLFs is shown as
$L_{\rm D} = 10^{22.9} \times (1+z)^{0.06\times z + 3.05}\ {\rm W}\ {\rm Hz}^{-1}$,
which can be used to select powerful radio AGNs at different redshifts 
through solely radio surveys (Section \ref{sec:turnoverL}).
This result also indicate a decreasing contribution of AGNs to entire radio populations towards higher redshift.
  \item The probability of a galaxy hosting a radio-excess AGN is shown as a function of 
stellar mass, radio luminosity, and redshift:
$p(L_{\rm R}\ |\ M_{\star},\ z) 
= 10^{-0.79} \left(\frac{M_{\star}}{M_0}\right)^{1.06} \left(\frac{L_{\rm R}}{L_0}\right)^{-0.77} \left(\frac{1+z}{1+z_0}\right)^{3.08}$ for SFGs,
$p(L_{\rm R}\ |\ M_{\star},\ z) 
= 10^{-0.70} \left(\frac{M_{\star}}{M_0}\right)^{1.41} \left(\frac{L_{\rm R}}{L_0}\right)^{-0.60} \left(\frac{1+z}{1+z_0}\right)^{2.47}$ for QGs (Section \ref{sec:pzevl}).
It indicates that radio-excess AGNs in QGs prefer to reside in more massive galaxies with
larger radio luminosity than those in SFGs.
The fractions of radio-excess AGNs in both SFGs and QGs are increasing
from the local universe to the higher redshift.
In addition, this increasing trend in SFGs is more significant than in QGs.
\end{enumerate}

The above studies can lay the foundation for further investigation
with the upcoming revolutionary radio facilities, 
such as Square Kilometer Array \citep[SKA;][]{Dewdney2009,Norris2013,McAlpine2015} 
and the Next Generation Very Large Array \citep[ngVLA;][]{Hughes2015}.

\begin{acknowledgements}

We thank the referee for constructive comments that greatly improved this paper.
This work is supported by the National Natural Science Foundation of China (Project No. 12173017 and Key Project No. 12141301).

\end{acknowledgements}

\bibliographystyle{aa}
\bibliography{ms.bib}

\onecolumn
\appendix
\renewcommand\thefigure{\thesection.\arabic{figure}}
\section{Spectral energy distribution (SED) fitting}
\setcounter{figure}{0}
\label{sec:SED}
\textsc{cigale} uses a series of AGN/galaxies templates to efficiently model the 
observed multi-wavelength data from the X-rays and far-ultraviolet (FUV) to
the far-infrared (FIR) and radio bands.
Through the Bayesian-like approach, this code estimates many fundamental physical properties,
such as star formation rate, stellar mass, dust luminosity, AGN contribution and other quantities.
In the fitting process, the galaxy templates used in this work includes the following six modules:
the star formation history, the single stellar population \citep{Bruzual2003}, 
the dust attenuation \citep{Calzetti2000}, the dust emission \citep{Dale2014},
and the AGN module \citep{Stalevski2012,Stalevski2016}.
These modules and their parameters are summarized in Table \ref{tab:SEDpara}.
In Fig. \ref{fig:SED}, we present the best-fit SEDs of four 
radio sources in the GOODS-N field as examples.

\begin{figure*}[h!]
\center
\begin{subfigure}[b]{0.4\linewidth}
\includegraphics[width=\linewidth, clip]{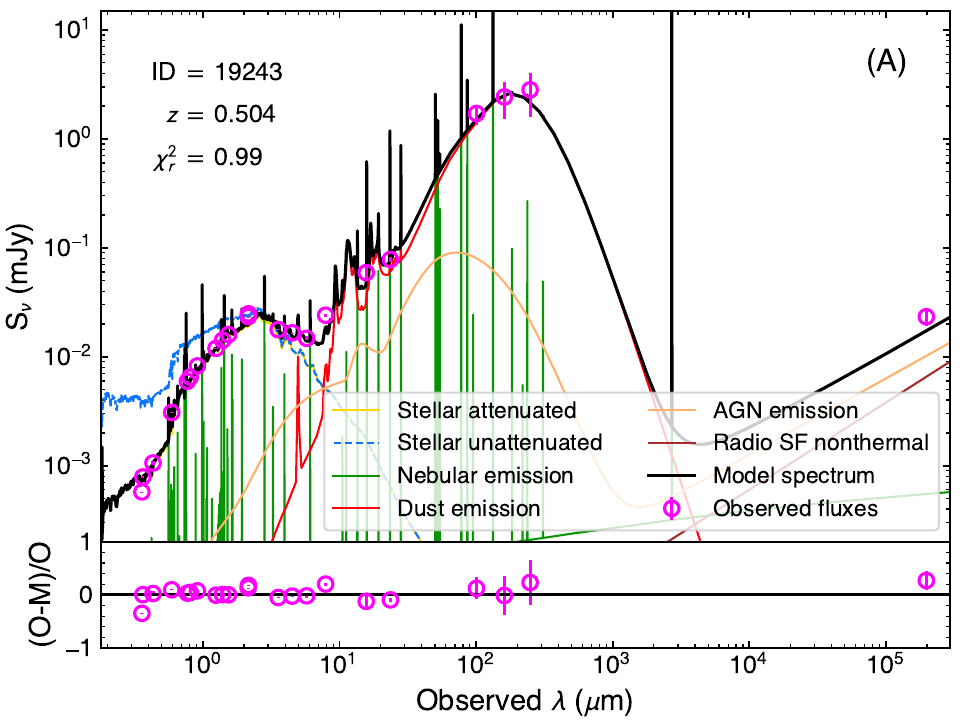}
\end{subfigure}
\begin{subfigure}[b]{0.4\linewidth}
\includegraphics[width=\linewidth, clip]{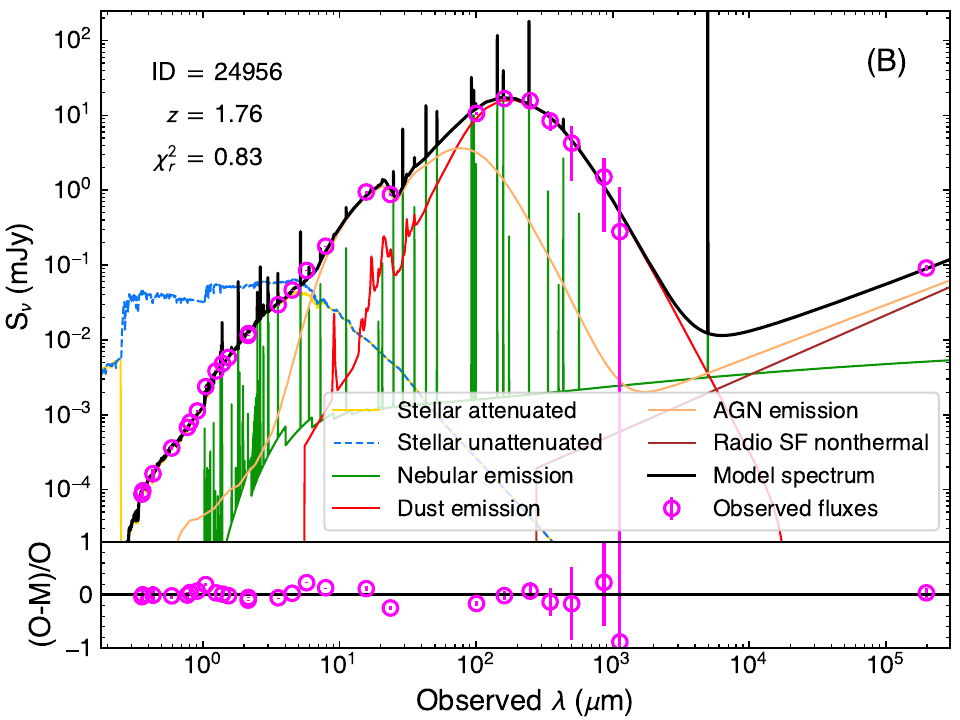}
\end{subfigure}
\begin{subfigure}[b]{0.4\linewidth}
\includegraphics[width=\linewidth, clip]{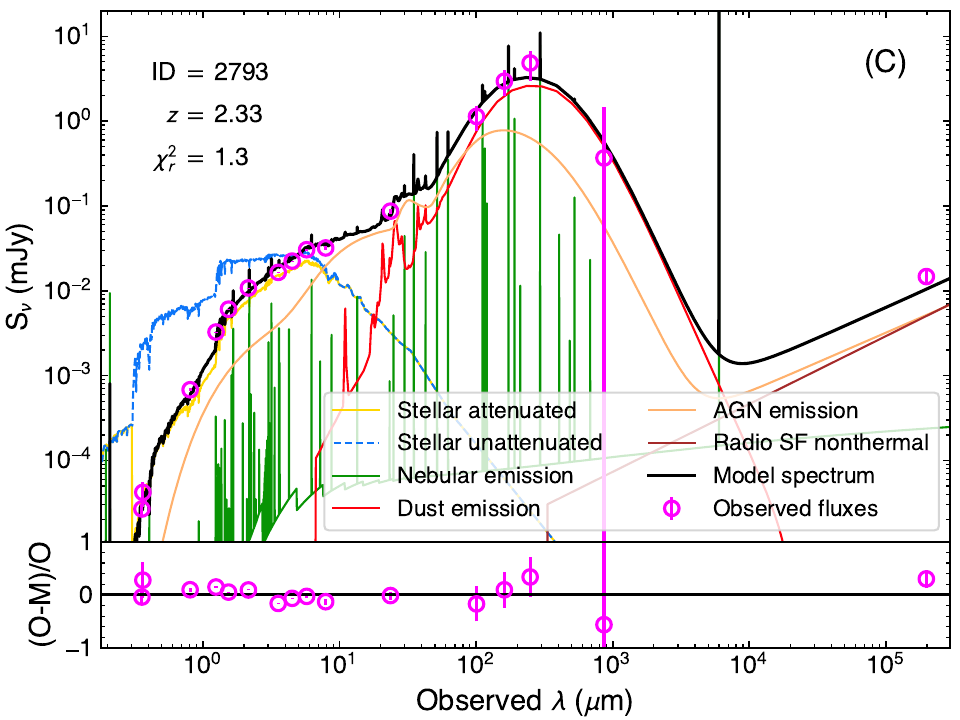}
\end{subfigure}
\begin{subfigure}[b]{0.4\linewidth}
\includegraphics[width=\linewidth, clip]{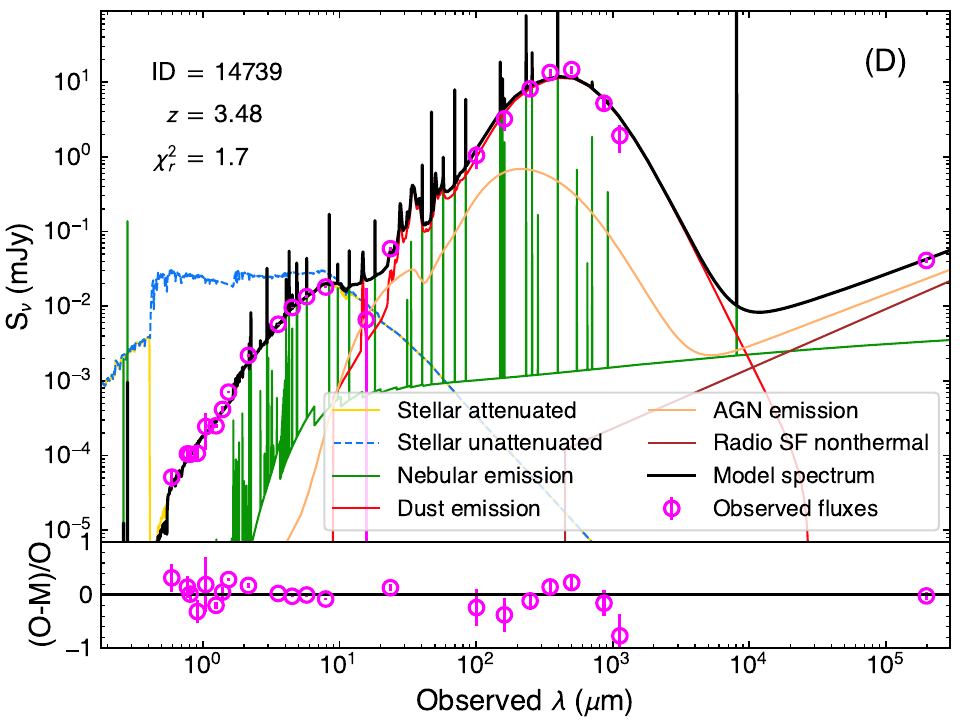}
\end{subfigure}
\caption{The best-fit SED model from \textsc{cigale} fitting for four radio sources as examples.
Panels (A), (B), (C), and (D) show radio sources from the GOODS-N field with redshift $z$ of 0.504, 1.76, 2.33, and 3.48, respectively.
\label{fig:SED}}
\end{figure*}

   \begin{table*}[h!]
   \caption{Model parameters for SED fitting with \textsc{cigale} \label{tab:SEDpara}}
   \centering
   \small
   \begin{tabular}{clcccccc}
   \hline\hline\xrowht[()]{10pt}
   Module & Parameter & Symbol & Values \\
   \hline\xrowht[()]{5pt}
Star formation history           & Stellar $e$-folding time & $\tau_{\rm{star}}$ ($10^6$ yr) & 50, 100, 250, 500, 1000, 2500, 5000, 10000 \\ 
$[\rm{SFR} \propto t \exp(-t/\tau)]$ & Stellar age                     & $t_{\rm{star}}$ ($10^6$ yr)     & 100, 250, 500, 1000, 2500, 5000, 10000 \\
  \hline\xrowht[()]{5pt}
Single stellar population      & Initial mass function      & --                                              & \cite{Chabrier2003} \\
 \citep{Bruzual2003}               & Metallicity                     & $Z$                                           & 0.02 \\
  \hline\xrowht[()]{5pt}
Dust attenuation  & \multirow{2}{*}{Color excess of the nebular lines} & \multirow{2}{*}{$E(B-V)$ (mag)}        & \multirow{2}{*}{0.005, 0.05, 0.1--0.7 (step 0.1), 0.9, 1.1, 1.3, 1.5} \\
\citep{Calzetti2000} & & &   \\ 
\hline\xrowht[()]{5pt}
Galactic dust emission & \multirow{2}{*}{Slope in $dM_{\rm{dust}} \propto U^{-\alpha}dU$} & \multirow{2}{*}{$\alpha$} & \multirow{2}{*}{1.25, 1.5, 1.75, 2.0, 2.25, 2.5, 2.75} \\
\citep{Dale2014} & & & \\
\hline\xrowht[()]{5pt}
\multirow{3}{*}{AGN (UV-to-IR)} & AGN contribution to IR luminosity & $\rm{frac}_{\rm{AGN}}$ & 0.0, 0.01, 0.1, 0.3, 0.5, 0.7, 0.9, 0.99 \\ 
          & Viewing angle & $\theta$ & $30^{\circ}$, $70^{\circ}$ \\
  & Polar-dust color excess & $E(B-V)_{\rm{PD}}$ (mag) & 0.0, 0.2, 0.4 \\
\hline\xrowht[()]{5pt}
\multirow{2}{*}{Radio} & SF radio-IR correlation parameter & $q_{\rm{IR}}$ & 2.58 \\
  & Radio-loudness parameter & $R_{\rm{AGN}}$ & 0.0, 0.01, 0.1, 1, 10, 100, 1000, 10000 \\
\hline
   \end{tabular}
   \end{table*}

\section{Correct the classification purity for the radio-excess AGN}
\label{sec:REAGNpurity}
\begin{figure*}[h!]
\includegraphics[width=\linewidth, clip]{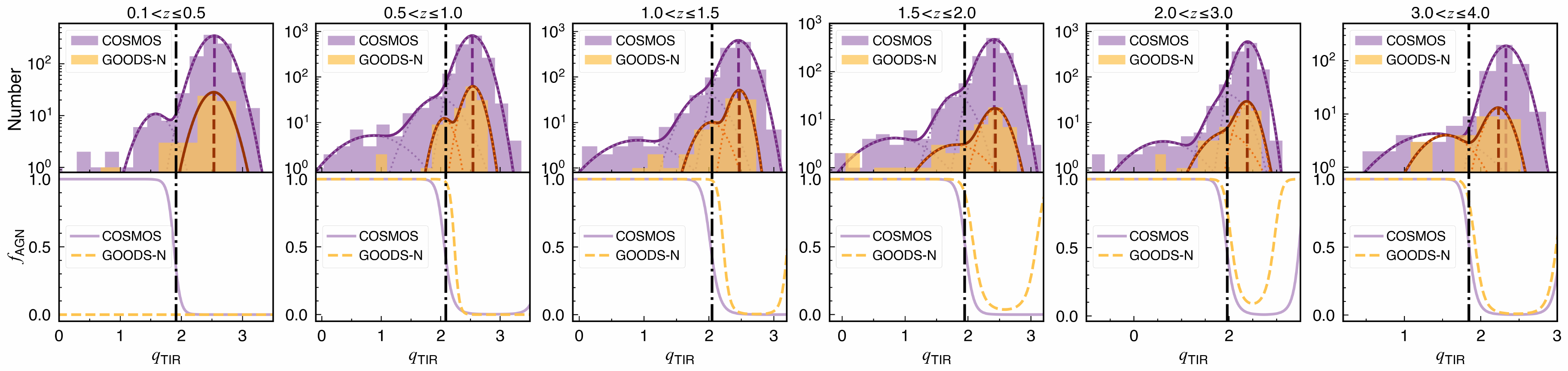}
\caption{
{\it Upper row:} Distribution of IR-radio-ratio ($q_{\rm TIR}$) for SFGs
in the GOODS-N (yellow histogram) 
and COSMOS/UltraVISTA (violet histogram) fields at different redshift bins (from left to right columns).
Each dotted curve represents each single gaussian model,
while the solid curve represents the best-fit model to the entire $q_{\rm TIR}$ distribution.
The vertical dashed line corresponds to the peak position of the highest gaussian component.
The vertical black dash-dotted line represents the threshold 
used in this work to separate SFGs and radio-excess AGNs 
(corresponds to $q_{\rm AGN}$ in Fig. \ref{fig:qIRdis} and see details in Section \ref{sec:qTIRagn}).
{\it Bottom row:} Fraction of radio-excess AGNs in all the radio objects 
for the GOODS-N (yellow dashed curve) 
and COSMOS field/UltraVISTA (violet dotted curve) fields.
Parameter $f_{\rm AGN}$ is defined as
$1-N_{\rm SFG}/N_{\rm TOT}$,
where $N_{\rm SFG}$ represents the height of the highest gaussian model 
(the highest dotted curve) at a certain $q_{\rm TIR}$ value,
and $N_{\rm TOT}$ represents the height of the entire best-fit model 
(entire solid curve) at a certain $q_{\rm TIR}$ value.
The vertical black dash-dotted line represents the threshold used in this work
to separate SFGs and radio-excess AGNs (same as the upper panels).
\label{fig:qIRfield}}
\end{figure*}

We use the cross point between the highest and the second highest gaussian components of $q_{\rm TIR}$ distribution
as the threshold to select radio-excess AGN (see details in Section \ref{sec:qTIRagn}).
Radio-excess AGNs with $q_{\rm TIR}$ around the selection threshold 
(corresponds to the $q_{\rm AGN}$ in Fig. \ref{fig:qIRdis} and the vertical black dash-dotted lines in Fig. \ref{fig:qIRfield})
are partially contaminated by the SFGs.
To correct the classification purity for radio-excess AGN,
we follow \cite{Delvecchio2022} to define $f_{\rm AGN}$ as the
$1-N_{\rm SFG}/N_{\rm TOT}$,
where $N_{\rm SFG}$ represents the best-fit gaussian model for the 
highest $q_{\rm TIR}$ peak (see the highest dotted curves in Fig. \ref{fig:qIRfield}),
and $N_{\rm TOT}$ represents the best-fit model 
for the entire $q_{\rm TIR}$ distribution (see the entire purple solid curves in Fig. \ref{fig:qIRfield}).
At the threshold $q_{\rm AGN}$ (see the black dash-dotted line in Fig. \ref{fig:qIRfield}), 
$f_{\rm AGN}$ is usually below 50$\%$ for samples in the COSMOS field,
while $f_{\rm AGN}$ is around 100$\%$ for the GOODS-N field.
It indicates that the {\RAS} in the GOODS-N have a nearly 100\% AGN purity under the selection threshold used in this work.
Therefore, we only make AGN purity corrections for the objects in the COSMOS fields.
Following \cite{Delvecchio2022}, we use the $L_{\rm 1.4 GHz}$ threshold converted from the median
IR luminosity in each redshift bin to make the purity corrections.

\newpage
\section{Radio luminosity function in each field: GOODS-N and COSMOS/UltraVISTA fields}
We obtain the radio luminosity function in each field following the same procedure described in Section \ref{sec:RLF}.
For both SFGs and radio-excess AGNs, RLFs in these three fields present generally consistent results 
(see detailed discussions in Section \ref{sec:RLF}).

\begin{figure*}[h!]
\centering
\includegraphics[width=0.93\linewidth, clip]{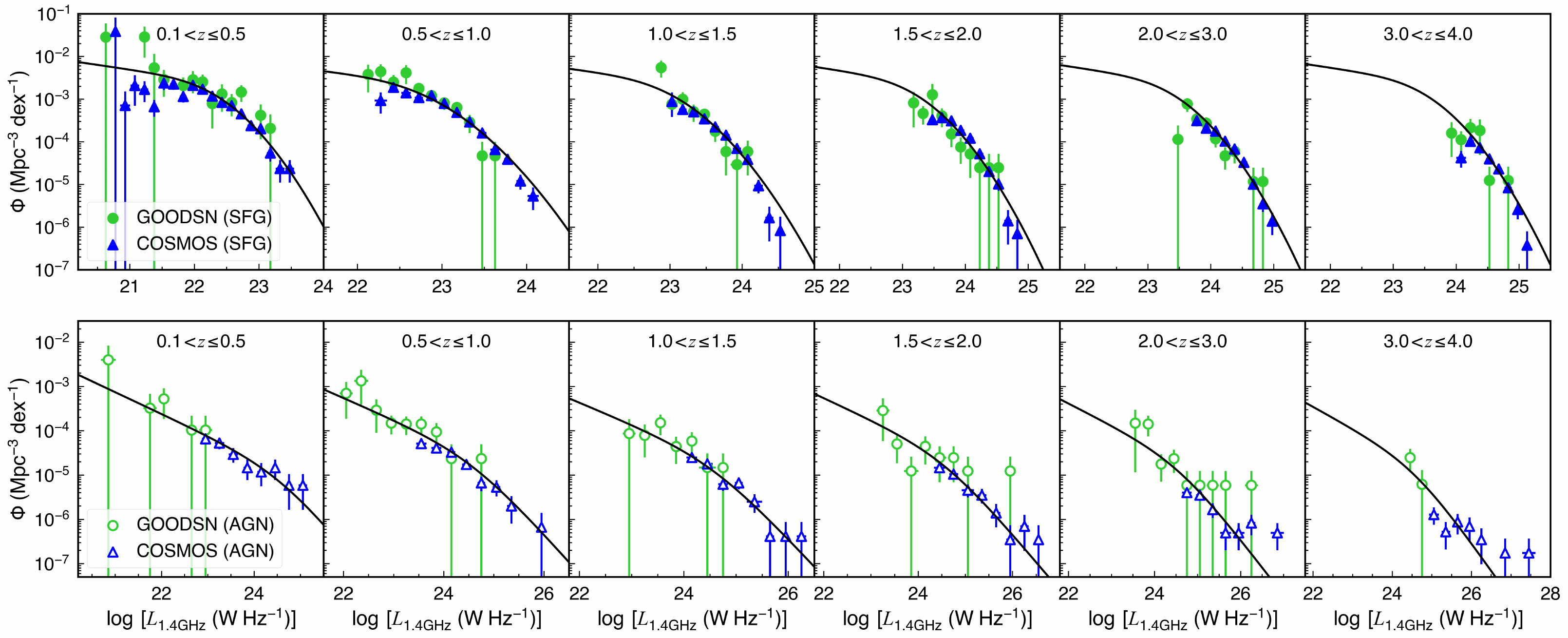}
\caption{The 1.4 GHz Radio luminosity function of SFGs ({\it top row}) 
and radio-excess AGNs ({\it bottom row}) in each field 
(green circles: GOODS-N; blue triangles: COSMOS/UltraVISTA).
The black solid curves in the {\it top row} represent the best-fit model (PLE) for SFGs.
The black solid curves in the {\it bottom row} represent the best-fit model (PDE) for radio-excess AGNs.
\label{fig:RLFfield}}
\end{figure*}

\newpage
\section{Tables of radio luminosity functions for SFGs and radio-excess AGNs}
\begin{table*}[h!]
\centering
\caption{Radio luminosity function of SFGs obtained with the $1/V_{\rm max}$ method. \label{tab:SFG}}
\begin{tabular}{ccc|ccc|ccc}
\hline\hline\xrowht[()]{10pt}
$\log L_{\rm 1.4 GHz}$ & $\log \Phi$ & $N$ & $\log L_{\rm 1.4 GHz}$ & $\log \Phi$ & $N$ & $\log L_{\rm 1.4 GHz}$ & $\log \Phi$ & $N$ \\
$[\rm{W\ Hz}^{-1}]$ & $[\rm{Mpc}^{-3}\ \rm{dex}^{-1}]$ & & $[\rm{W\ Hz}^{-1}]$ & $[\rm{Mpc}^{-3}\ \rm{dex}^{-1}]$ & & $[\rm{W\ Hz}^{-1}]$ & $[\rm{Mpc}^{-3}\ \rm{dex}^{-1}]$ & \\
\hline
\multicolumn{3}{c|}{$0.1<z\leq 0.5$} & \multicolumn{3}{c|}{$0.5<z\leq 1.0$} & \multicolumn{3}{c}{$1.0<z\leq 1.5$} \\
\hline
$21.07 \pm 0.05$ & $-2.68 \pm 0.28$ & 4 &  $22.12 \pm 0.03$ & $-2.42 \pm 0.26$ & 4 &  $22.88 \pm 0.03$ & $-2.26 \pm 0.17$ & 16 \\
$21.22 \pm 0.02$ & $-1.81 \pm 0.27$ & 12 &  $22.27 \pm 0.04$ & $-2.57 \pm 0.15$ & 14 &  $23.02 \pm 0.03$ & $-3.08 \pm 0.14$ & 43 \\
$21.38 \pm 0.02$ & $-3.18 \pm 0.16$ & 9 &  $22.43 \pm 0.02$ & $-2.66 \pm 0.10$ & 54 &  $23.18 \pm 0.02$ & $-3.10 \pm 0.11$ & 162 \\
$21.52 \pm 0.04$ & $-2.58 \pm 0.15$ & 28 &  $22.57 \pm 0.03$ & $-2.55 \pm 0.14$ & 135 &  $23.32 \pm 0.02$ & $-3.30 \pm 0.09$ & 282 \\
$21.68 \pm 0.04$ & $-2.65 \pm 0.10$ & 48 &  $22.72 \pm 0.03$ & $-2.84 \pm 0.06$ & 213 &  $23.47 \pm 0.03$ & $-3.40 \pm 0.07$ & 297 \\
$21.82 \pm 0.03$ & $-2.79 \pm 0.13$ & 45 &  $22.88 \pm 0.02$ & $-2.92 \pm 0.06$ & 332 &  $23.62 \pm 0.02$ & $-3.70 \pm 0.08$ & 222 \\
$21.97 \pm 0.02$ & $-2.60 \pm 0.13$ & 82 &  $23.02 \pm 0.02$ & $-3.09 \pm 0.06$ & 377 &  $23.77 \pm 0.02$ & $-3.99 \pm 0.09$ & 157 \\
$22.12 \pm 0.03$ & $-2.67 \pm 0.11$ & 124 &  $23.18 \pm 0.02$ & $-3.25 \pm 0.07$ & 299 &  $23.93 \pm 0.04$ & $-4.15 \pm 0.05$ & 79 \\
$22.27 \pm 0.03$ & $-3.01 \pm 0.13$ & 133 &  $23.32 \pm 0.03$ & $-3.54 \pm 0.09$ & 204 &  $24.07 \pm 0.03$ & $-4.31 \pm 0.19$ & 46 \\
$22.43 \pm 0.03$ & $-2.96 \pm 0.16$ & 125 &  $23.47 \pm 0.04$ & $-3.79 \pm 0.04$ & 115 &  $24.22 \pm 0.05$ & $-5.03 \pm 0.13$ & 11 \\
$22.57 \pm 0.03$ & $-3.11 \pm 0.12$ & 111 &  $23.62 \pm 0.04$ & $-4.18 \pm 0.06$ & 47 &  $24.38 \pm 0.04$ & $-5.78 \pm 0.31$ & 2 \\
$22.72 \pm 0.03$ & $-3.02 \pm 0.13$ & 80 &  $23.77 \pm 0.03$ & $-4.41 \pm 0.08$ & 29 &   & & \\
$22.88 \pm 0.04$ & $-3.62 \pm 0.07$ & 39 &  $23.93 \pm 0.03$ & $-4.92 \pm 0.14$ & 9 &   & & \\
$23.02 \pm 0.02$ & $-3.51 \pm 0.21$ & 35 &  $24.07 \pm 0.05$ & $-5.27 \pm 0.22$ & 4 &   & & \\
$23.18 \pm 0.04$ & $-4.26 \pm 0.14$ & 9 &   & & &  & & \\
$23.32 \pm 0.05$ & $-4.63 \pm 0.22$ & 4 &   & & &  & & \\
$23.47 \pm 0.05$ & $-4.63 \pm 0.22$ & 4 &   & & &  & & \\
\hline
\multicolumn{3}{c|}{$1.5<z\leq 2.0$} & \multicolumn{3}{c|}{$2.0<z\leq 3.0$} & \multicolumn{3}{c}{$3.0<z\leq 4.0$} \\
\hline
$23.18 \pm 0.06$ & $-3.09 \pm 0.31$ & 2 &  $23.62 \pm 0.05$ & $-3.11 \pm 0.13$ & 13 &  $23.93 \pm 0.04$ & $-3.80 \pm 0.31$ & 2 \\
$23.32 \pm 0.05$ & $-3.34 \pm 0.18$ & 6 &  $23.77 \pm 0.02$ & $-3.48 \pm 0.07$ & 96 &  $24.07 \pm 0.03$ & $-4.11 \pm 0.17$ & 12 \\
$23.47 \pm 0.03$ & $-3.10 \pm 0.24$ & 73 &  $23.93 \pm 0.03$ & $-3.62 \pm 0.08$ & 211 &  $24.22 \pm 0.03$ & $-3.80 \pm 0.16$ & 81 \\
$23.62 \pm 0.02$ & $-3.41 \pm 0.10$ & 230 &  $24.07 \pm 0.03$ & $-3.83 \pm 0.06$ & 317 &  $24.38 \pm 0.04$ & $-3.89 \pm 0.22$ & 120 \\
$23.77 \pm 0.03$ & $-3.63 \pm 0.07$ & 283 &  $24.22 \pm 0.03$ & $-4.12 \pm 0.07$ & 226 &  $24.52 \pm 0.03$ & $-4.40 \pm 0.05$ & 81 \\
$23.93 \pm 0.03$ & $-3.88 \pm 0.08$ & 208 &  $24.38 \pm 0.03$ & $-4.19 \pm 0.09$ & 176 &  $24.68 \pm 0.04$ & $-4.63 \pm 0.06$ & 58 \\
$24.07 \pm 0.03$ & $-4.06 \pm 0.09$ & 162 &  $24.52 \pm 0.04$ & $-4.48 \pm 0.05$ & 91 &  $24.82 \pm 0.04$ & $-5.07 \pm 0.09$ & 22 \\
$24.22 \pm 0.04$ & $-4.28 \pm 0.05$ & 70 &  $24.68 \pm 0.04$ & $-4.99 \pm 0.08$ & 30 &  $24.97 \pm 0.03$ & $-5.58 \pm 0.16$ & 7 \\
$24.38 \pm 0.03$ & $-4.70 \pm 0.08$ & 27 &  $24.82 \pm 0.03$ & $-5.45 \pm 0.14$ & 10 &   & & \\
$24.52 \pm 0.03$ & $-4.99 \pm 0.12$ & 14 &  $24.97 \pm 0.03$ & $-5.86 \pm 0.22$ & 4 &   & & \\
\hline
\end{tabular}
\tablefoot{
$N$ is the source number in each luminosity bin.
}
\end{table*}

\newpage
\begin{table*}[h!]
\centering
\caption{Radio luminosity function of radio-excess AGNs obtained with the $1/V_{\rm max}$ method. \label{tab:AGN}}
\begin{tabular}{ccc|ccc|ccc}
\hline\hline\xrowht[()]{10pt}
$\log L_{\rm 1.4 GHz}$ & $\log \Phi$ & $N$ & $\log L_{\rm 1.4 GHz}$ & $\log \Phi$ & $N$ & $\log L_{\rm 1.4 GHz}$ & $\log \Phi$ & $N$ \\
$[\rm{W\ Hz}^{-1}]$ & $[\rm{Mpc}^{-3}\ \rm{dex}^{-1}]$ & & $[\rm{W\ Hz}^{-1}]$ & $[\rm{Mpc}^{-3}\ \rm{dex}^{-1}]$ & & $[\rm{W\ Hz}^{-1}]$ & $[\rm{Mpc}^{-3}\ \rm{dex}^{-1}]$ & \\
\hline
\multicolumn{3}{c|}{$0.1<z\leq 0.5$} & \multicolumn{3}{c|}{$0.5<z\leq 1.0$} & \multicolumn{3}{c}{$1.0<z\leq 1.5$} \\
\hline
$22.05 \pm 0.05$ & $-3.28 \pm 0.27$ & 3 &  $22.05 \pm 0.11$ & $-3.15 \pm 0.31$ & 2 &  $23.25 \pm 0.07$ & $-4.10 \pm 0.29$ & 3 \\
$22.95 \pm 0.06$ & $-4.18 \pm 0.09$ & 12 &  $22.35 \pm 0.12$ & $-2.87 \pm 0.30$ & 3 &  $23.55 \pm 0.05$ & $-3.82 \pm 0.20$ & 19 \\
$23.25 \pm 0.08$ & $-4.27 \pm 0.10$ & 24 &  $22.65 \pm 0.03$ & $-3.53 \pm 0.29$ & 3 &  $23.85 \pm 0.09$ & $-4.35 \pm 0.25$ & 53 \\
$23.55 \pm 0.08$ & $-4.53 \pm 0.14$ & 22 &  $22.95 \pm 0.09$ & $-3.83 \pm 0.18$ & 15 &  $24.15 \pm 0.05$ & $-4.38 \pm 0.15$ & 62 \\
$23.85 \pm 0.06$ & $-4.83 \pm 0.19$ & 18 &  $23.25 \pm 0.06$ & $-3.85 \pm 0.18$ & 45 &  $24.45 \pm 0.08$ & $-4.74 \pm 0.07$ & 43 \\
$24.15 \pm 0.06$ & $-4.93 \pm 0.22$ & 11 &  $23.55 \pm 0.05$ & $-4.01 \pm 0.13$ & 81 &  $24.75 \pm 0.07$ & $-5.20 \pm 0.11$ & 16 \\
$24.45 \pm 0.03$ & $-4.83 \pm 0.19$ & 6 &  $23.85 \pm 0.05$ & $-4.17 \pm 0.15$ & 65 &  $25.05 \pm 0.06$ & $-5.17 \pm 0.11$ & 16 \\
$24.75 \pm 0.05$ & $-5.23 \pm 0.31$ & 4 &  $24.15 \pm 0.06$ & $-4.48 \pm 0.06$ & 49 &  $25.35 \pm 0.13$ & $-5.60 \pm 0.18$ & 6 \\
$25.05 \pm 0.07$ & $-5.23 \pm 0.31$ & 5 &  $24.45 \pm 0.07$ & $-4.76 \pm 0.09$ & 26 &   & & \\
 & & & $24.75 \pm 0.05$ & $-5.17 \pm 0.14$ & 10 &   & & \\
 & & & $25.05 \pm 0.06$ & $-5.27 \pm 0.15$ & 8 &   & & \\
 & & & $25.35 \pm 0.07$ & $-5.70 \pm 0.25$ & 3 &   & & \\
\hline
\multicolumn{3}{c|}{$1.5<z\leq 2.0$} & \multicolumn{3}{c|}{$2.0<z\leq 3.0$} & \multicolumn{3}{c}{$3.0<z\leq 4.0$} \\
\hline
$23.25 \pm 0.12$ & $-3.54 \pm 0.34$ & 3 &  $23.55 \pm 0.08$ & $-3.83 \pm 0.40$ & 2 &  $24.45 \pm 0.07$ & $-4.60 \pm 0.22$ & 4 \\
$23.55 \pm 0.08$ & $-4.29 \pm 0.27$ & 3 &  $23.85 \pm 0.07$ & $-3.84 \pm 0.20$ & 9 &  $25.05 \pm 0.10$ & $-5.89 \pm 0.16$ & 4 \\
$24.15 \pm 0.01$ & $-4.35 \pm 0.26$ & 8 &  $24.15 \pm 0.04$ & $-4.75 \pm 0.25$ & 8 &  $25.35 \pm 0.08$ & $-6.28 \pm 0.25$ & 7 \\
$24.45 \pm 0.08$ & $-4.70 \pm 0.19$ & 30 &  $24.45 \pm 0.07$ & $-4.63 \pm 0.22$ & 23 &  $25.65 \pm 0.06$ & $-6.06 \pm 0.19$ & 3 \\
$24.75 \pm 0.04$ & $-4.75 \pm 0.22$ & 42 &  $24.75 \pm 0.09$ & $-5.39 \pm 0.09$ & 23 &  $25.95 \pm 0.03$ & $-6.15 \pm 0.22$ & 5 \\
$25.05 \pm 0.11$ & $-5.34 \pm 0.12$ & 33 &  $25.05 \pm 0.06$ & $-5.45 \pm 0.09$ & 21 &  $26.25 \pm 0.05$ & $-6.45 \pm 0.31$ & 4 \\
$25.35 \pm 0.07$ & $-5.45 \pm 0.14$ & 13 &  $25.35 \pm 0.05$ & $-5.78 \pm 0.14$ & 10 &   & & \\
$25.65 \pm 0.07$ & $-5.85 \pm 0.22$ & 10 &  $25.65 \pm 0.10$ & $-6.30 \pm 0.25$ & 3 &   & & \\
$26.25 \pm 0.07$ & $-6.15 \pm 0.31$ & 4 &  $25.95 \pm 0.11$ & $-6.30 \pm 0.25$ & 3 &   & & \\
 & & & $26.25 \pm 0.07$ & $-6.08 \pm 0.19$ & 5 &   & & \\
 & & & $26.85 \pm 0.13$ & $-6.30 \pm 0.25$ & 3 &   & & \\
\hline
\end{tabular}
\tablefoot{
$N$ is the source number in each luminosity bin.
}
\end{table*}

\newpage

\section{Radio luminosity function for radio-excess AGN split into the subsets hosted by SFGs and QGs}

We obtain the radio luminosity function for radio-excess AGN hosted by different populations (SFGs and QGs)
following the same procedure described in Section \ref{sec:RLF}.
We found a similar evolution trend for radio-excess AGN hosted by different populations 
to that for low-excitation radio galaxies in \cite{Kondapally2022} (see detailed discussions in Section \ref{sec:RLFAGN}).

\begin{figure*}[h!]
\centering
\includegraphics[width=0.75\linewidth, clip]{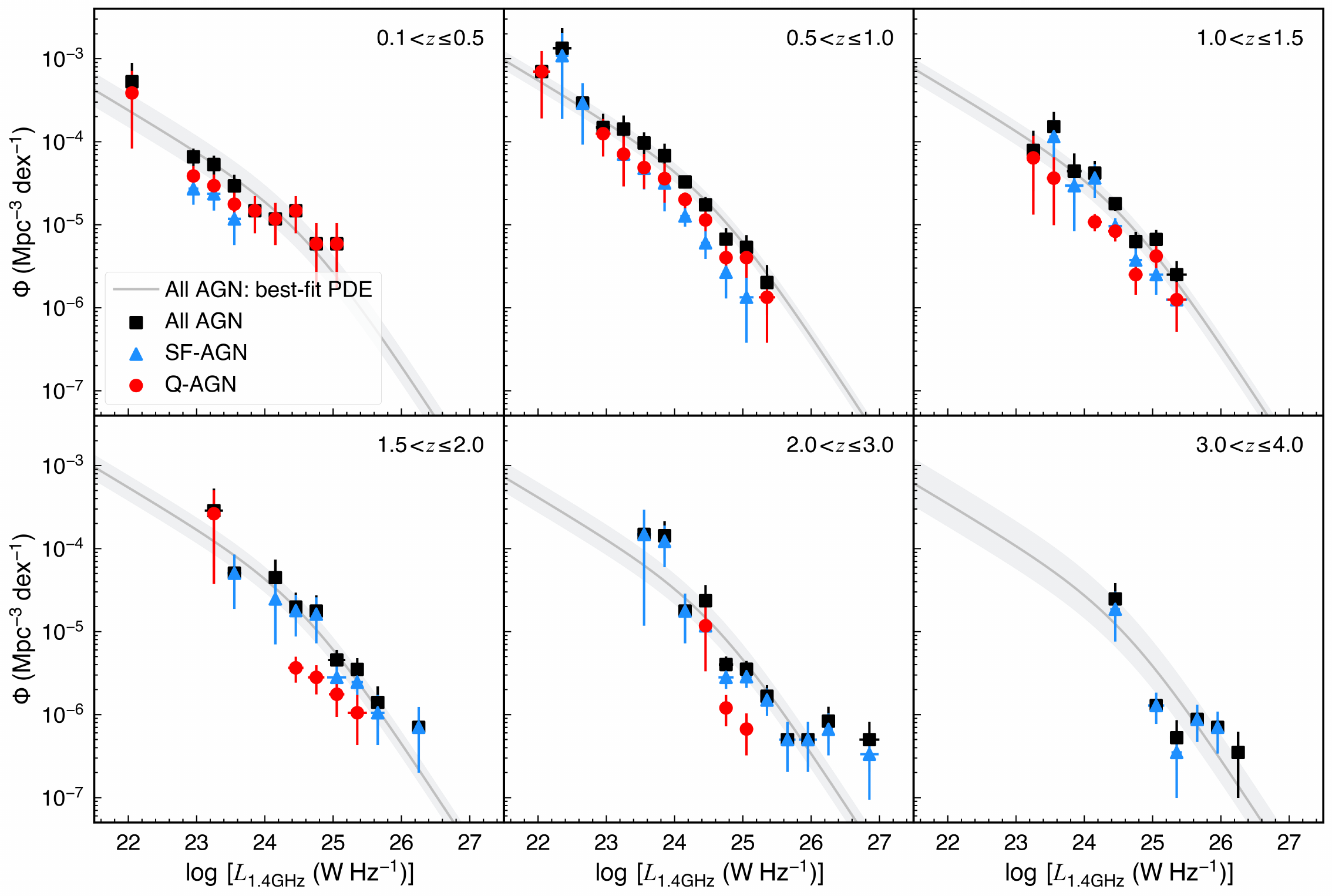}
\caption{The 1.4 GHz Radio luminosity function of
all the radio-excess AGNs (black squares) split into the subsets hosted by SFGs (blue triangles) and QGs (red circles).
The grey solid curves represent the best-fit PDE model for all the radio-excess AGNs
and the grey regions represent the 1-$\sigma$ uncertainty of the best-fit model.
\label{fig:RLFSFGQG}}
\end{figure*}

\twocolumn
\section{Simple $\chi^2$ fits to the probability of hosting a radio-excess AGN in each redshift bin}
\label{sec:apxchi2}
\begin{figure}[h!]
\centering
\includegraphics[width=0.83\linewidth, clip]{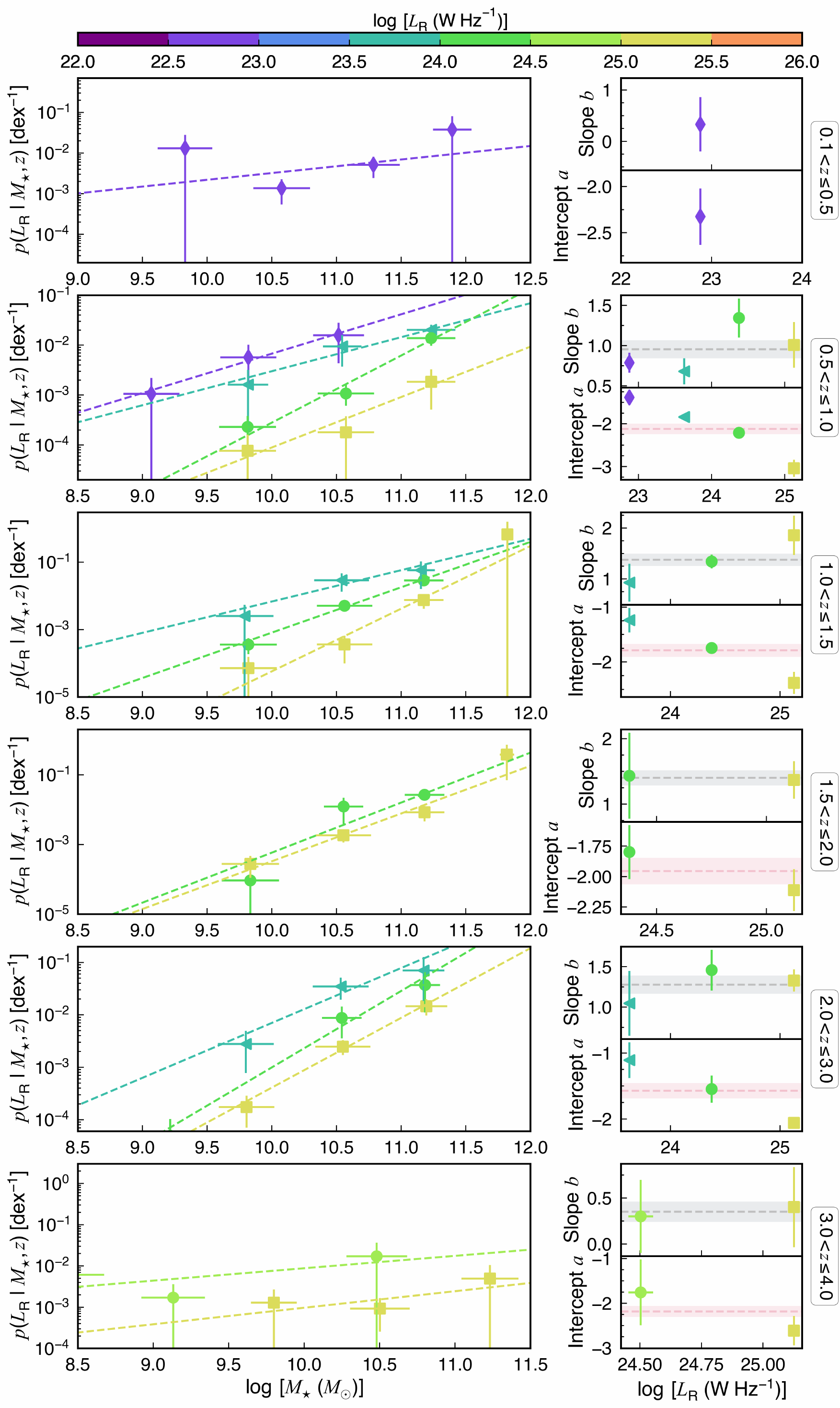}
\caption{{\it Left column:} The probability density function $p(L_{\rm R}\ |\ M_{\star},\ z)$
of a SFG hosting a radio-excess AGN with a given 1.4 GHz luminosity $L_{\rm R}$ 
as a function of stellar mass $M_{\star}$ at different redshift bins (from upper to bottom rows).
Different colors represent different $L_{\rm R}$ bins.
Each dashed line represents the best-fit power-law relation for $p(L_{\rm R}\ |\ M_{\star},\ z)$ 
as a function of $M_{\star}$
in each $L_{\rm R}$ bin (defined by Equation \ref{eq:FagnLR}).
{\it Right column:} Best-fit slope $b$ (upper panel at each redshift bin) 
and intercept $a$ (bottom panel at each redshift bin) of the above power-law relation.
Different colors represent different $L_{\rm R}$ bins (same as those in the left column).
In the upper-right panel of each redshift bin, the gray dashed line 
shows the mean of slopes over all the $L_{\rm R}$ bins ($\bar{b}$),
while the gray region represents $\bar{b} \pm 0.1$.
In the bottom-right panel of each redshift bin, the pink dashed line 
shows the mean of intercepts over all the $L_{\rm R}$ bins ($\bar{a}$),
while the pink region represents $\bar{a} \pm 0.1$.
The region scale ($\pm 0.1$) is just a reference value for 
showing the difference between each data point with the mean value. 
\label{fig:FagnSFGdatawithmass}}
\end{figure}

\begin{figure}[h!]
\centering
\includegraphics[width=0.83\linewidth, clip]{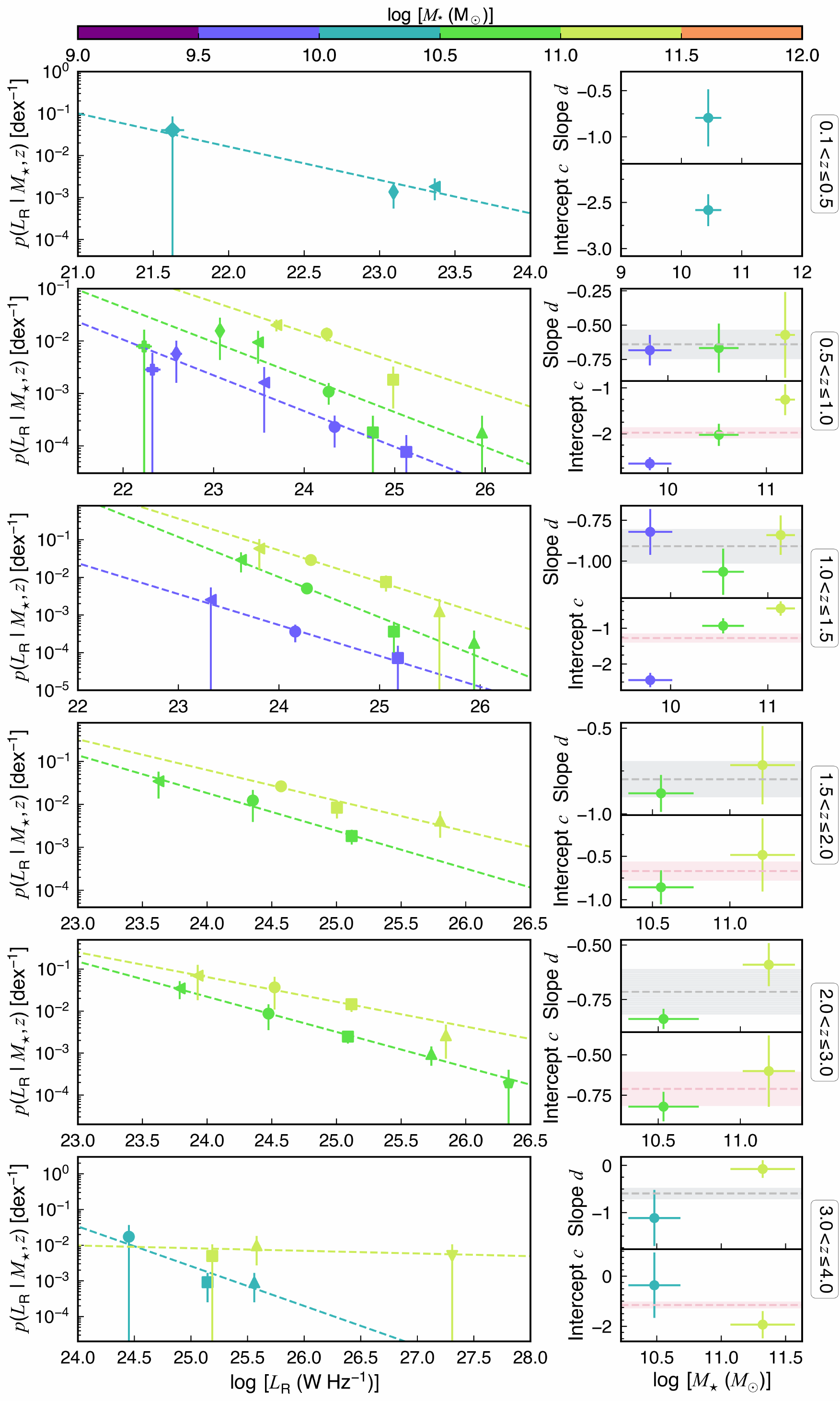}
\caption{{\it Left column:} The probability density function $p(L_{\rm R}\ |\ M_{\star},\ z)$
of a SFG with a given stellar mass $M_{\star}$ hosting a radio-excess AGN 
as a function of 1.4 GHz luminosity $L_{\rm R}$ at different redshift bins (from upper to bottom rows).
Different colors represent different $M_{\star}$ bins.
Each dashed line represents the best-fit power-law relation for $p(L_{\rm R}\ |\ M_{\star},\ z)$ 
as a function of $L_{\rm R}$
in each $M_{\star}$ bin (defined by Equation \ref{eq:Fagnmass}).
{\it Right column:} Best-fit slope $d$ (upper panel at each redshift bin) 
and intercept $c$ (bottom panel at each redshift bin) of the above power-law relation.
Different colors represent different $M_{\star}$ bins (same as the left column).
In the upper-right panel of each redshift bin, the gray dashed line 
shows the mean of slopes over all the $M_{\star}$ bins ($\bar{d}$),
while the gray region represents $\bar{d} \pm 0.1$.
In the bottom-right panel of each redshift bin, the pink dashed line 
show the mean of intercepts over all the $M_{\star}$ bins ($\bar{c}$),
while the pink region represents $\bar{c} \pm 0.1$.
The region scale ($\pm 0.1$) is just a reference value for 
showing the difference between each data point with the mean value. 
\label{fig:FagnSFGdatawithLR}}
\end{figure}

As Equation \ref{eq:FagnLR} or \ref{eq:Fagnmass} shows, at each fixed $L_{\rm R}$ (or $M_\star$),
we assume a simple power-law relation for $p(L_{\rm R}\ |\ M_{\star},\ z)$ as
a function of $M_{\star}$ (or $L_{\rm R}$).
Then we apply $\chi^2$ fits to the data 
(see colored symbols in Fig. \ref{fig:FagnSFGdatawithmass} and Fig. \ref{fig:FagnSFGdatawithLR}).
The best-fit model in each $L_{\rm R}$ (or $M_\star$) bin
is shown as the colored line in the left column of 
Fig. \ref{fig:FagnSFGdatawithmass} (or Fig. \ref{fig:FagnSFGdatawithLR}).
The best-fit parameters in each $L_{\rm R}$ (or $M_\star$) bin 
are shown as colored points in the right column of 
Fig. \ref{fig:FagnSFGdatawithmass} (or Fig. \ref{fig:FagnSFGdatawithLR}).
For more details about analysis and discussion, we refer readers to Section \ref{sec:maximumz}.

\onecolumn
\section{The probability of hosting a radio-excess AGN in QGs and all the galaxies}
Similar to Section \ref{sec:pzevl} (mainly for SFGs), here we show the probability of hosting a radio-excess AGN in QGs and all the galaxies based on maximum-likelihood fitting over the entire redshift range.

\begin{figure*}[h!]
\centering
\includegraphics[width=0.95\linewidth, clip]{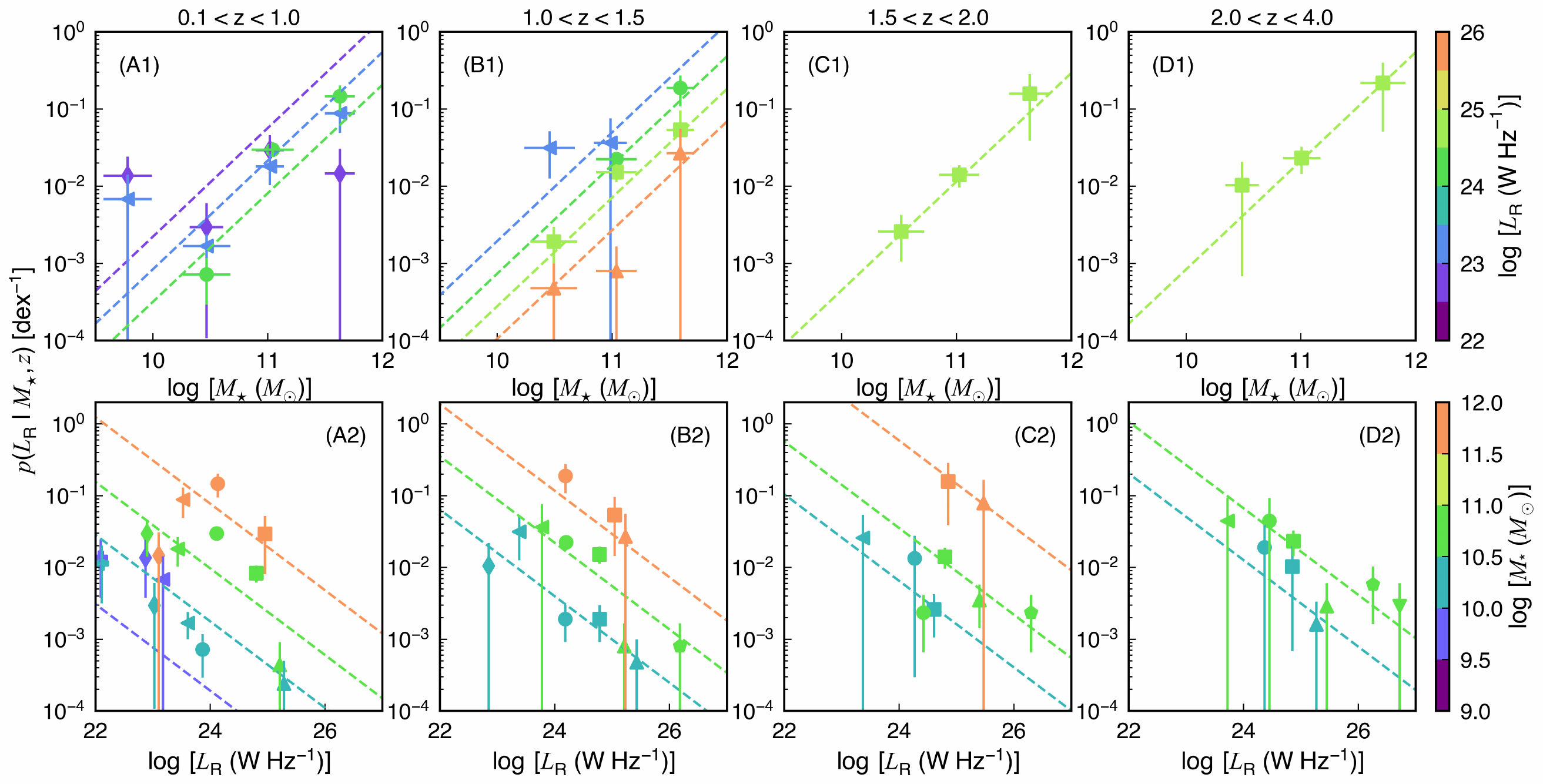}
\caption{Estimates of $p(L_{\rm R}\ |\ M_{\star},\ z)$
as a function of $M_{\star}$ and $L_{\rm R}$
based on our maximum-likelihood fitting results for QGs.
Colored dashed lines in both top and bottom rows
represent our best-fit model from the unbinned maximum-likelihood fitting (see Equation \ref{eq:pLMz} and Table \ref{tab:Fagnz})
through combining data in the three fields (GOODS-N and COSMOS/UltraVISTA)
over the entire redshift range ($0.1 < z < 4$) evaluated at the center of each redshift bin.
Binned data points are scaled with the probability estimated by the model (see Equation \ref{eq:pLMz})
using the $N_{\rm obs}/N_{\rm mdl}$ method of \cite{Aird2012} (see details in Section \ref{sec:maximumz}).
In the top panel, different colors represent different $M_{\star}$ bins.
In the bottom panel, different colors indicate different $L_{\rm R}$ bins. 
\label{fig:FagnQGmo}}
\end{figure*}

\begin{figure*}[h!]
\centering
\includegraphics[width=0.95\linewidth, clip]{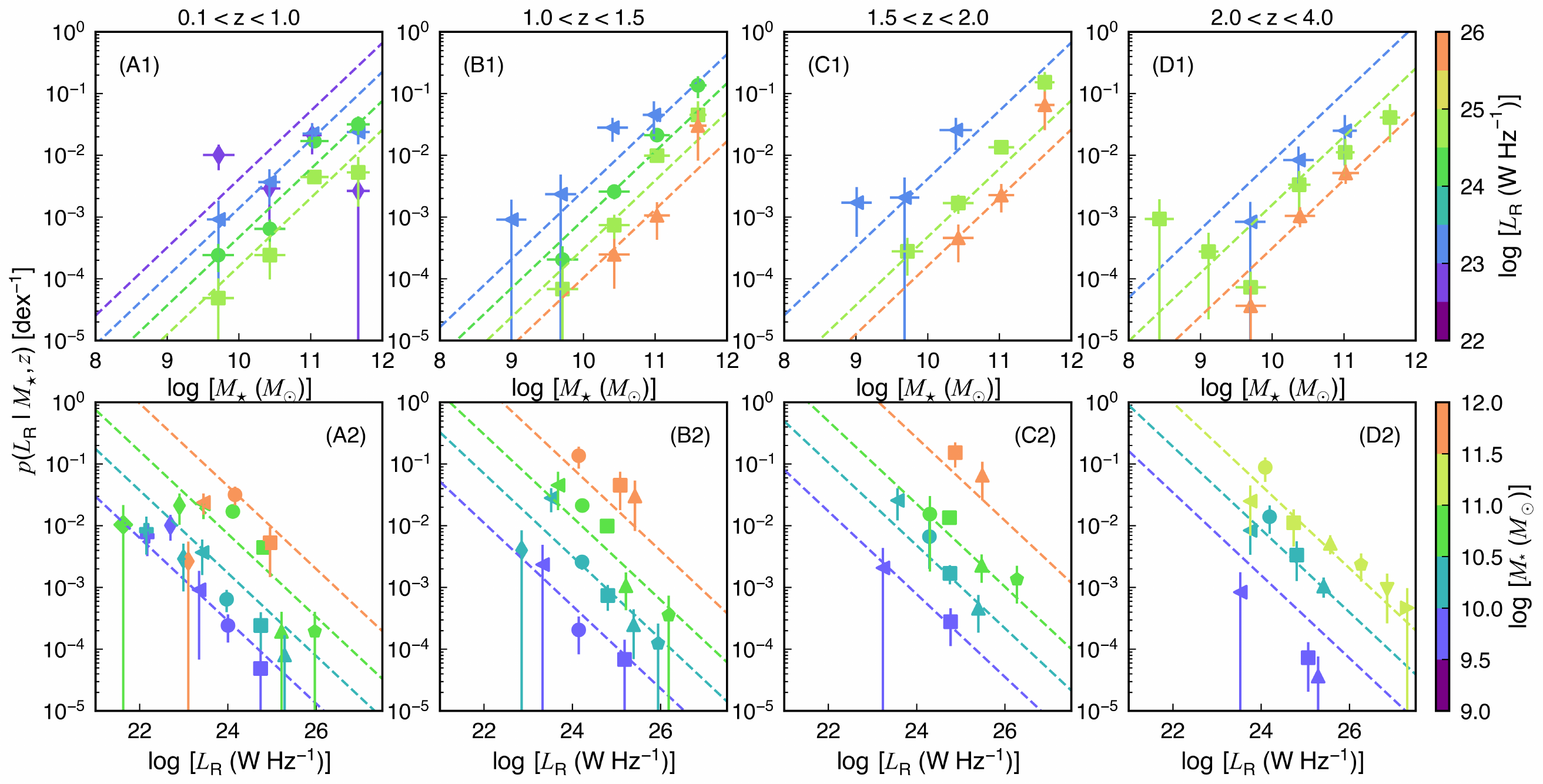}
\caption{Same as Fig. \ref{fig:FagnSFGmo} and Fig. \ref{fig:FagnQGmo} but for all galaxies. 
\label{fig:FagnTotmo}}
\end{figure*}

\twocolumn
\section{The evolution of $p(L_{\rm R}\ |\ M_{\star},\ z)$ in SFGs considering the $M_\star$-dependent IRRC}
\begin{figure}[h!]
\includegraphics[width=0.8\linewidth, clip]{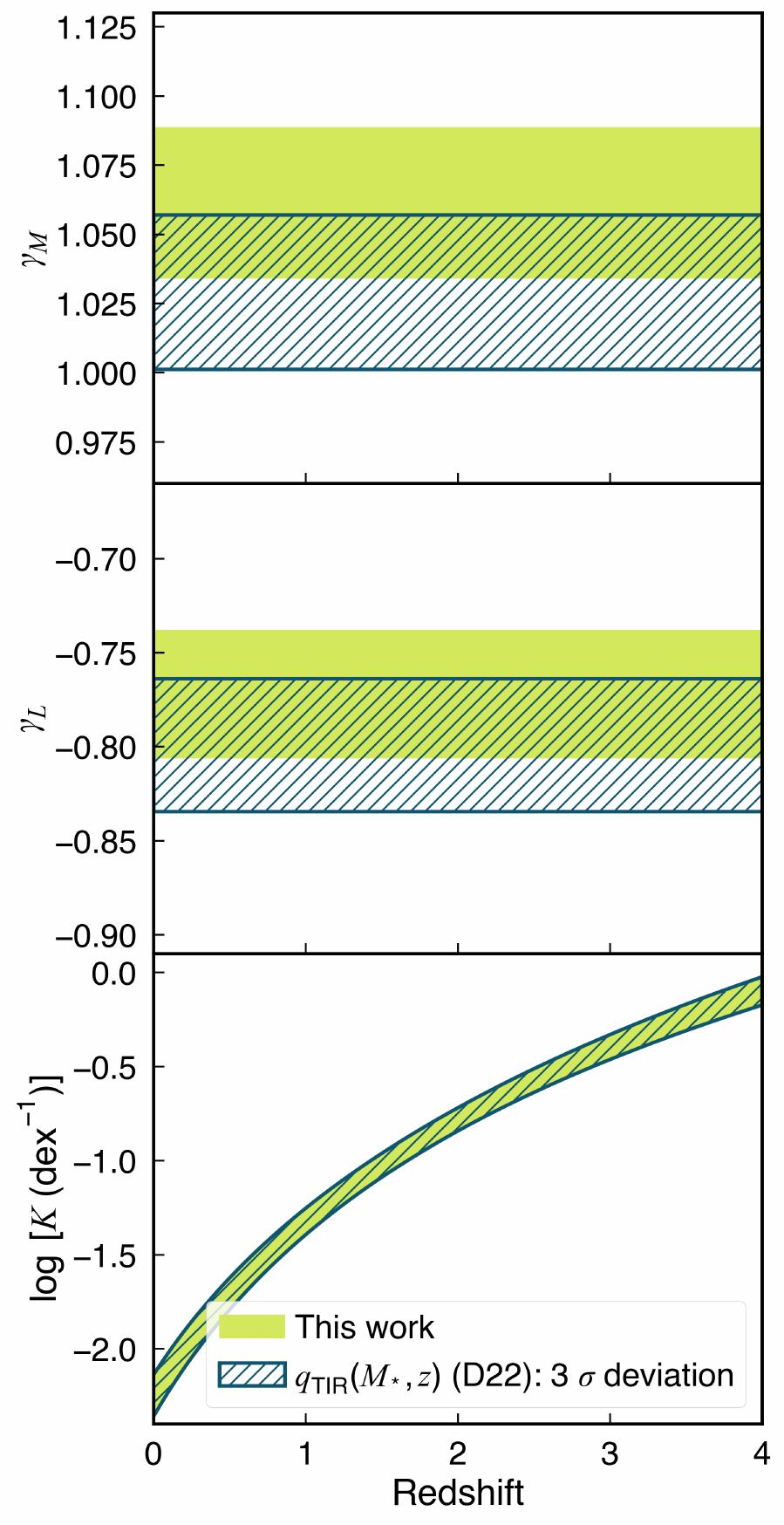}
\caption{
Best-fit parameters of the maximum-likelihood fitting for 
$p(L_{\rm R}\ |\ M_{\star},\ z)$ in SFGs over the entire redshift range.
Green regions correspond to the best-fit parameters
based on the sample selected by the method in Section \ref{sec:REAGNpaper} 
(same as the blue region in Fig. \ref{fig:Fagnevo}).
Dark blue regions correspond to the best-fit parameters
based on the sample selected by the $M_*$-dependent IRRC 
\cite[][see details in Section \ref{sec:Mqir}]{Delvecchio2021}.
\label{fig:Fagnevocompare}}
\end{figure}

In order to examine whether the $M_\star$-dependent IRRC will affect our results,
we follow \cite{vanderVlugt2022} to select radio-excess AGNs
with $q_{\rm TIR}$ deviating more than $3\sigma$
from the $M_\star$-dependent IRRC of \cite{Delvecchio2021}. 
To compare with the results shown in Section \ref{sec:pzevl}
(see the green region in Fig. \ref{fig:Fagnevocompare}),
we also use Equation \ref{eq:pLMz} to recalculate the probability of SFGs hosting a radio-excess AGN
(see the dark blue region in Fig. \ref{fig:Fagnevocompare}).
Fig. \ref{fig:Fagnevocompare} shows that using the 
$M_*$-dependent IRRC \citep{Delvecchio2021} does not alter the results in this work.

\end{CJK*}
\end{document}